\documentclass[10pt]{article}
\usepackage[accepted]{tmlr} 

\usepackage{amsmath,amsfonts,bm}

\def\eqref#1{equation~\ref{#1}}

\def\1{\bm{1}}

\DeclareMathAlphabet{\mathsfit}{\encodingdefault}{\sfdefault}{m}{sl}
\SetMathAlphabet{\mathsfit}{bold}{\encodingdefault}{\sfdefault}{bx}{n}

\newcommand{\E}{\mathbb{E}}

\newcommand{\Var}{\mathrm{Var}}

\newcommand{\Cov}{\mathrm{Cov}}

\usepackage{tcolorbox}
\tcbuselibrary{listings, breakable, skins}
\definecolor{iclrgray}{HTML}{F5F5F5}
\definecolor{iclrblue}{HTML}{D0EFFF}
\definecolor{iclrteal}{HTML}{94d2bd}
\definecolor{iclrgreen}{HTML}{99d98c}
\definecolor{softteal}{RGB}{0,150,150}
\definecolor{softblue}{RGB}{70,130,180}

\usepackage[utf8]{inputenc}
\usepackage[T1]{fontenc}
\usepackage{hyperref}
\usepackage{url}
\usepackage{booktabs}
\usepackage{amsfonts}
\usepackage{nicefrac}
\usepackage{microtype}
\usepackage{xcolor}

\usepackage{pgffor}

\usepackage{dsfont}
\usepackage{graphicx}
\usepackage{caption}
\usepackage{babel}

\usepackage{wrapfig}
\usepackage{tabularx}

\usepackage{amsmath}
\usepackage{amssymb}
\usepackage{mathtools}
\usepackage{amsthm}
\usepackage{float}
\usepackage{subcaption}

\usepackage{algorithm,algcompatible}

\usepackage{tikz}
\usetikzlibrary{positioning}
\usetikzlibrary{shapes.geometric, arrows}

\tikzstyle{server} = [rectangle, draw, fill=blue!10, text centered, minimum height=2em]
\tikzstyle{user} = [circle, draw, fill=green!10, text centered, minimum size=2em]
\tikzstyle{arrow} = [thick,->,>=stealth]
\tikzstyle{data} = [rectangle, draw, dashed, fill=yellow!10, text centered, minimum height=2em]
\tikzstyle{attacker} = [circle, draw, fill=red!10, text centered, minimum size=2em]

\usepackage{array}
\usepackage[capitalize,noabbrev]{cleveref}

\makeatletter
\let\@fnsymbol\@arabic
\makeatother

\newtheorem{theorem}{Theorem}[section]

\theoremstyle{definition}

\theoremstyle{remark}

\usepackage[suppress]{color-edits} 
\addauthor[suppress]{ae}{blue}
\addauthor{aek}{cyan}
\addauthor{elm}{blue}

\title{Approaching the Harm of Gradient Attacks While Only Flipping Labels}

\author{
\name Abdessamad El-Kabid \email \href{mailto:abdessamad.el-kabid@polytechnique.edu}{abdessamad.el-kabid@polytechnique.edu} \\
\addr École Polytechnique
\AND
\name El-Mahdi El-Mhamdi \email \href{mailto:el-mahdi.el-mhamdi@polytechnique.edu}{el-mahdi.el-mhamdi@polytechnique.edu}\\
\addr École Polytechnique
}

\def\month{08}  
\def\year{2026} 
\def\openreview{\url{https://openreview.net/forum?id=eAqBcljvsU}} 

\begin{document}

\maketitle

\begin{abstract}
Machine learning systems deployed in distributed or federated environments are highly susceptible to adversarial manipulations, particularly availability attacks -- rendering the trained model unavailable.
Prior research in distributed ML has demonstrated such adversarial effects through the injection of gradients or data poisoning.  In this work, we ask whether comparable degradation is still possible under a substantially more constrained action space: the adversary may only flip a limited number of labels of existing training examples, without modifying features, injecting samples, or directly controlling gradients. 
We analyze the extent of damage caused by constrained label flipping attacks against distributed learning under mean aggregation -- the dominant baseline in research and production.
Focusing on classification problems, (1) we propose a novel formalization of label flipping attacks as a per-round constrained optimization problem, derive a greedy label-selection rule for logistic regression,
and empirically evaluate it beyond its derivation setting, including on MLPs and robust aggregators. 
The rule is provably per-epoch optimal for the attacker under the mean aggregator.
(2) Empirically, we show that optimized label flipping can cause substantial accuracy degradation while outperforming random label flipping under similar budgets.
(3) We shed light on an interesting interplay between what the attacker gains from more \emph{write-access} versus what they gain from more \emph{flipping budget}.
(4) \aeedit{Finally, although the attack is derived for mean aggregation, we find that it can transfer empirically to 
the coordinate-wise median and trimmed mean aggregators, where its effectiveness  approaches that of the Little-is-Enough gradient attack. This demonstrates that even highly constrained label-flipping adversaries can pose a significant availability threat to distributed learning.}


\end{abstract}


\newcommand{\problem}[1]{{\color{red} /!$\backslash$ : #1}}
\newcommand{\Mahdi}[1]{{\color{blue} EM:#1}}
\newcommand{\Abdessamad}[1]{{\color{orange} Ab : #1}}

\section{Introduction and Related Work}

\begin{wrapfigure}{r}{0.5\textwidth}
\vspace{-5\baselineskip}
    \centering
    \begin{tikzpicture}[scale=0.7,font=\small]


\draw[
  thick,
  gray,
  dashed
]
(0.5,-0.3) circle ({sqrt(3)});

\filldraw[
  thick,
  draw=teal,
  dotted,
  fill=teal!20,
  rotate=15
] 
(0.8,-1) circle (1);

\filldraw[
  thick,
  draw=magenta!70!black,
  dash dot,
  fill=magenta!20,
]
(1.45,-1.05) circle (0.5);

\filldraw[
  line width=1.2pt,
  draw=green!70!black,
  fill=green!30
]
(1.65,-1.25) circle (0.2);

\node[gray] at (-0.5,1.5) {\large $\mathbb{R}^d$};


\begin{scope}[xshift=2.7cm, yshift=0.5cm]

  \draw[thick,gray,dashed] (0,0) -- (0.7,0);
  \node[anchor=west] at (0.7,0) {$\mathbb{R}^d$ };
  
  \draw[thick,teal,dotted] (0,-0.5) -- (0.7,-0.5);
  \node[anchor=west] at (0.7,-0.5) {$\nabla_{\theta}L\bigl(h_{\theta}(\mathcal{X}), \mathcal{Y}\bigr)$};
  
  \draw[thick,magenta!70!black,dash dot] (0,-1.0) -- (0.7,-1.0);
  \node[anchor=west] at (0.7,-1.0) {$\nabla_{\theta}L\bigl(h_{\theta}(F_{\mathcal{X}}),F_{\mathcal{Y}}\bigr)$};
  
  \draw[line width=1.2pt,green!70!black] (0,-1.5) -- (0.7,-1.5);
  \node[anchor=west] at (0.7,-1.5) {$\nabla_{\theta}L\bigl(\mathcal{D}_{\mathcal{X}},h_{\theta}(F_{\mathcal{Y}})\bigr)$};

\end{scope}

\end{tikzpicture}
    \caption{Images of the gradient operator on different sets. $\mathbb{R}^d$ is where an attacker can craft unrestricted gradient attacks. Given a model $h_{\theta}$, $\nabla_{\theta} L(h_{\theta}(\mathcal{X}), \mathcal{Y})$ is the set of possible gradients given an unrestricted data poisoning \cite{bouaziz2024invertinggradientattacksmakes}, $\nabla_{\theta} L(h_{\theta}(\mathcal{F}_{\mathcal{X}}), \mathcal{F}_{\mathcal{Y}})$ is the set of possible gradients when data poisoning is restricted to a feasible set $\mathcal{F}_{\mathcal{X}} \times \mathcal{F}_{\mathcal{Y}} \subseteq \mathcal{X} \times \mathcal{Y}$, and $\nabla_{\theta}L\bigl(\mathcal{D}_{\mathcal{X}},h_{\theta}(F_{\mathcal{Y}})\bigr)$ is the set of possible gradients when the features are restricted to those in the dataset $\mathcal{D}_{\mathcal{X}}$ and the labels are chosen in the set of feasible labels.}
    \label{fig:image-grad}
    \vspace{-2\baselineskip}
\end{wrapfigure}

\label{sec:intro}
Machine learning systems can become prime targets for adversarial attacks.  \emph{Training-phase poisoning attacks} in particular have gained considerable attention as the widespread use of machine learning in critical applications has grown (\cite{10.1145/3006384,zhang2017understandingdeeplearningrequires,paudice2018labelsanitizationlabelflipping,Lu2022IndiscriminateDP,10.1145/2046684.2046692}). In these attacks, an adversary manipulates the training data in order to degrade or control the final trained model. Unlike evasion and backdoor attacks, poisoning requires no control over inference-time input: it suffices to manipulate part of the training set. Among such attacks, \emph{label flipping} stands out for its simplicity: The attacker simply changes the class label of a subset of training points while leaving other aspects of the data intact. This is especially relevant in federated learning situations where features are fixed by upstream data pipelines and the set of possible labels is predefined. For example, a common scenario is when workers are asked to label predefined images, the labels being in a finite set.  We address the following question in the context of a classification problem:  \begin{center}
    \aekedit{\emph{Can a gradient-informed attacker severely degrade a model using only label flips on existing data, under budget constraints ?}}
\end{center} 
In this work, we provide a positive answer to this question. By viewing label flipping as a constrained optimization problem from the point of view of the attacker, we show that carefully selected flips can steer the aggregated gradient away from its honest direction and reduce accuracy or render training unstable - even when only \(5\%\) of the labels are altered.


\begin{figure}[ht]
    \centering
    \includegraphics[width=0.8\linewidth]{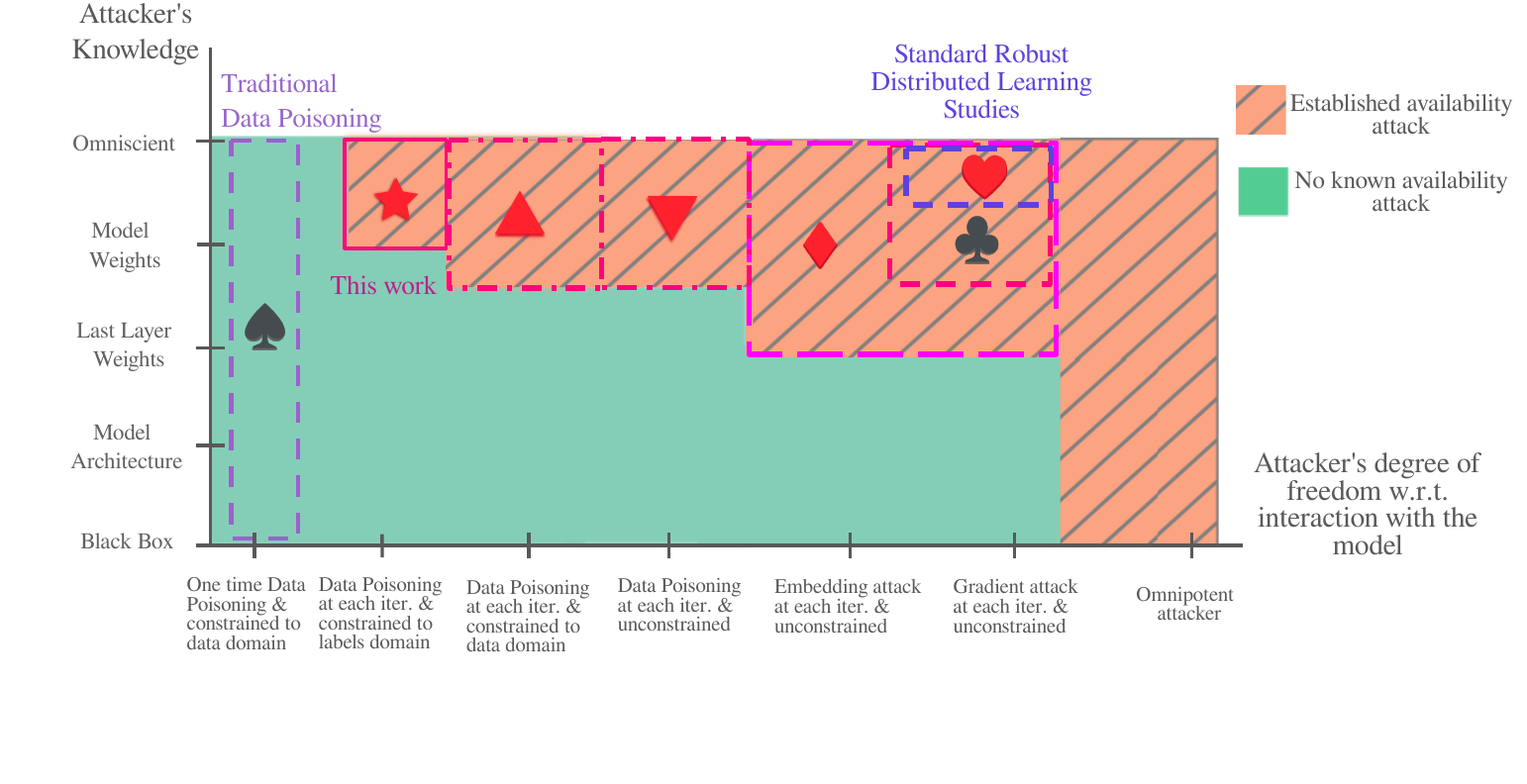}
    \caption{Territory of known availability attacks (in orange) within a domain of constraints. The closer to the origin, the more constrained is the setting for the attacker and the harder it is to realize an availability attack.
    $\spadesuit$: \cite{geiping2021witches, zhao_clpa_2022,ning2021invisible, huang2020metapoison}, 
    $\heartsuit$: \cite{blanchard2017byzantinetolerant,baruch2019little}, 
    $\clubsuit$: \cite{mhamdi2018hidden},
    $\diamondsuit$ so far only in convex settings : \cite{farhadkhani2022equivalence},
    $\triangle \& \triangledown$ : \cite{bouaziz2024invertinggradientattacksmakes},
    $\bigstar$ : Our contribution.
    }
    \label{fig:this-work}
\end{figure}


Previous work has investigated poisoning attacks and granted the adversary far more leverage than some realistic scenarios allow.
Some require the ability to modify \emph{both} labels and features
or even overwrite entire gradients (~\cite{bouaziz2024invertinggradientattacksmakes,baruch2019little});
others rely on injecting crafted examples into the training set (~\cite{shafahi2018poisonfrogstargetedcleanlabel,zhao_clpa_2022,koh_stronger_2021}).
\cite{Li_quinguru} proposes an agglomerative-clustering strategy to select vulnerable samples, assuming access to all training data and performing offline clustering. \cite{LablePoisoningIsAllYouNeed} demonstrates that carefully chosen label poisoning can insert backdoors via trajectory-matching methods. Their objective is to create targeted misclassification when a trigger is present, not to reduce overall model accuracy. Thus, although both their work and ours manipulate labels only, the goals, threat models, and evaluation metrics differ: their work aims to insert backdoors, whereas we study availability degradation under a tight budget and online setting. Likewise, \cite{Lavaur_systematic} applies random label flipping in the context of collaborative intrusion detection systems. The attack of \cite{Lavaur_systematic} targets domain-specific labels (e.g., DDoS) and does not formalize the attack as an optimization problem. In contrast, we aim to study an attack that is gradient-informed and can be proved to be optimal at each iteration for the considered objective. 

\cite{GNN_LF} study label-flipping attacks on GNNs under a federated learning setup. Their approach is oriented towards graph-structured data and relies on the unique properties of GNN aggregation mechanisms. In contrast, we propose a method that is formulated as a general optimization framework based on gradient alignment, making it applicable to a wider range of classification tasks.
When manipulation is limited to labels alone, existing methods typically
presuppose control over the vast majority of them (\(\ge 85\%\)) or access to the validation set, and require the training of a number of classifiers linear in the number of classes and the budget (~\cite{liu2023transferable,paudice2018labelsanitizationlabelflipping}), unsuitable for online or real-time training, especially at scale.

Another recent work, \citet{Bal2025}, focuses on {\em defense} in the binary classification setting via adversarial training (SVMs, untargeted), whereas our work formulates the {\em attacker’s} online, budget-constrained optimization and proves per-epoch greedy optimality under explicit write-access and local budget constraints.
Furthermore, our attack requires only control over a small fraction of data and changes no features at all, only labels. This expands the scope of known \emph{availability attacks} to a more constrained, yet practically damaging, threat model.

Following \cite{bouaziz2024invertinggradientattacksmakes}, Figure~\ref{fig:this-work} shows label-flipping attacks on the landscape of availability attacks, situating our contribution within the literature. Meanwhile, Figure~\ref{fig:image-grad} illustrates how label flipping compares to more general gradient-based attacks, particularly with respect to the set of gradients achievable under increasingly restrictive conditions.

\paragraph{Contributions.}
Our contributions are as follows. First, we formalize label flipping for logistic regression as a budget-constrained optimization problem for distributed classification under mean aggregation (Section~\ref{sec:model}). Second, we derive a greedy label-selection rule and prove that it is \emph{optimal for the attacker at each training step} under the mean-aggregation objective (Section~\ref{sec:Multiclass_case}). Experiments on standard \aeedit{scaled} image classification benchmarks show that optimized \aekedit{gradient-informed} label flipping causes substantially stronger availability degradation than random label flipping under matched budgets (Section~\ref{sec:multiclass_results}), with the added bonus of bypassing typical \aeedit{norm-based} anomaly detection thresholds (Appendix~\ref{appdx:grad_norms}). 
We also discuss how the attack depends on the decomposition of the global corruption rate $\rho = kb$, showing that broader write access $k$ can be more damaging than a larger local flipping budget $b$.
\aeedit{We further provide empirical evidence that the attack transfers beyond mean aggregation to coordinate-wise median and trimmed-mean aggregations, where constrained greedy label flipping approaches in effectiveness the celebrated Little-is-Enough gradient attack.}

\section{General Setting}
\label{sec:model}

\subsection{Notation}
\label{sec:notation}

\begin{table}[htb]
  \centering
  \caption{Notation Summary}
  \label{tab:notation}
  \begin{tabularx}{\textwidth}{|c|X|}
    \hline
    \textbf{Notation} & \textbf{Description} \\ \hline
    $t$ & Training-round, epoch, or iteration index. \\ \hline
    $x_i\in\mathbb{R}^{d+1}$ & Feature vector of example $i$, including a bias coordinate. \\ \hline
    $y_i^H\in\mathcal Y$ & Clean label of example $i$ before poisoning. \\ \hline
    $\tilde y_i\in\mathcal Y$ & Label submitted by the adversary after possible flipping. \\ \hline
    $\theta_t$ & Model parameter at round $t$. For binary logistic regression, $\theta_t=\alpha_t\in\mathbb{R}^{d+1}$; for multi-class softmax regression, $\theta_t=W_t\in\mathbb{R}^{C\times(d+1)}$. \\ \hline
    $H_t$ & Honest examples at round $t$, whose labels are not controlled by the adversary. \\ \hline
    $K_t^H$ & Clean version of the examples controlled by the adversary before label flipping. \\ \hline
    $K_t(\tilde y)$ & Attacker-controlled examples after label flipping. \\ \hline
    $D_t^H=H_t\cup K_t^H$ & Clean batch before poisoning. \\ \hline
    $D_t(\tilde y)=H_t\cup K_t(\tilde y)$ & Batch observed by the server after poisoning. \\ \hline
    $N=|D_t^H|=|D_t(\tilde y)|$ & Batch size. \\ \hline
    $k_t=|K_t^H|/|D_t^H|$ & Attacker's write-access fraction at round $t$. When fixed across rounds, we write $k$. \\ \hline
    $b\in[0,1]$ & Local flipping budget: the fraction of labels in $K_t^H$ that may be changed at round $t$. \\ \hline
    $\rho=kb$ & Global corruption budget: the maximum fraction of the full batch whose labels are flipped. \\ \hline
    $g_t^H$ & Clean gradient that would be computed on $D_t^H$. \\ \hline
    $g_t(\tilde y)$ & Poisoned gradient computed on $D_t(\tilde y)$. \\ \hline
    $\Delta_t$ & Reference direction used by the adversary in the directional label-flipping objective. \\ \hline
    $\mathds{1}[\cdot]$ & Indicator function. \\ \hline
  \end{tabularx}
\end{table}

The main notation used throughout the paper is summarized in Table~\ref{tab:notation}. Although the batch, the model parameters, and the attack all vary with the epoch $t$, we omit the epoch index whenever there is no risk of ambiguity. This is because the attacker's optimization problem is treated epoch-wise: at each epoch, the attacker observes the current training state, chooses a feasible subset of labels to flip, and the server then performs its update using the resulting poisoned batch.

We use $i\in D_t$ and $(x_i,y_i)\in D_t$ interchangeably when referring to data points by index. Similarly, writing $y_i\in K_t$ refers to the labels of the data points contained in $K_t$.


\subsection{Threat Model}

\paragraph{Rationale.}
As noted in~\cite{bouaziz2024invertinggradientattacksmakes}, the fundamental difference between gradient attacks and data poisoning comes from the limited expressivity of the latter. A gradient attacker may directly craft an arbitrary vector in parameter space, while a data-poisoning attacker can only influence the model through gradients induced by feasible data points. This difference is even more pronounced for label-flipping.

Our goal is to understand how much availability damage remains possible under this strongly restricted action space. We therefore study an adversary that can only flip labels of existing data, under a strict per-epoch budget. The attacker cannot modify features, inject new examples, delete honest data, directly overwrite gradients, or change the server-side optimizer. \aekedit{In terms of knowledge, the attacker is assumed to have read access to the gradients, an assumption commonly used in the gradient attack litterature~\cite{blanchard2017byzantinetolerant, el2021distributed, yin2021byzantinerobust,lilisu2018distribstatml}.}
In this sense, the attacker is weaker than classical gradient or model-poisoning adversaries in terms of write actions.

Following the two works most closely related to ours~\cite{farhadkhani2022equivalence, bouaziz2024invertinggradientattacksmakes}, we also allow the malicious worker to recalculate its attack at each iteration, similarly to Algorithm~1 in~\cite{steinhardt2017certified}.

\begin{wrapfigure}{r}{0.5\textwidth}
  \centering
   \vspace{-2\baselineskip}
  \begin{tikzpicture}[node distance=2cm, align=center]
    \tikzstyle{server} = [rectangle, draw, fill=blue!10, text centered, minimum height=2em]
    \tikzstyle{user}   = [circle, draw, fill=green!10, text centered, minimum size=2em]
    \tikzstyle{arrow}  = [thick,->,>=stealth]
    \tikzstyle{data}   = [rectangle, draw, dashed, fill=yellow!10, text centered, minimum height=2em]
    \tikzstyle{attacker} = [circle, draw, fill=red!10, text centered, minimum size=2em]

    \foreach \i in {1,...,4} {
        \node (worker\i) [user] at (\i*1.7cm-3cm, 0) {Worker};
    }
    \node (batch)   [data, above of=worker3]    {Training batch $D$};
    \node (server)  [server, above of=batch]    {Server};
    \node (reservoir) [data, below of=worker3]  {Clean Data Reservoir};
    \node (attacker) [attacker] at (5*1.7cm-3cm, 0){Attacker};

    \draw[arrow] (batch) -- (server);
    \foreach \i in {1,...,4} {
        \draw[arrow] (worker\i) -- (batch) node[midway, above] {$I_\i$};
    }
    \draw[arrow] (attacker) -- (batch) node[midway, above] {$K_t(\tilde y)$};
    \foreach \i in {1,...,4} {
        \draw[arrow, dashed] (reservoir) -- (worker\i) node[midway, below] {$I_\i$};
    }
    \draw[arrow, dashed] (reservoir) -- (attacker) node[midway, below] {$K_t^H$};
  \end{tikzpicture}
  \caption{Illustration of the setting: Each user obtains its data from a clean reservoir. The malicious user flips up to a budget \(b\) fraction of the labels in $K_t^H$.}
  \label{fig:Setting}
  \vspace{-1\baselineskip}
\end{wrapfigure}

\paragraph{General Learning Setup.} 
We study a classification problem where the training dataset is denoted by  
$
\mathcal{D}_{\text{train}} = \{(x_i, y_i)\}_{i=1}^n,
$
with samples drawn from a distribution \(\mathcal{D}\) over the input–label space \(\mathcal{X} \times \mathcal{Y}\). The learner trains a neural network \(h_\theta\), parameterized by \(\theta \in \mathbb{R}^d\), using an iterative optimization procedure to minimize a loss function \(\mathcal{L}\). The objective is to achieve low test loss on a held-out dataset \(\mathcal{D}_{\text{test}}\).

We model the learning process as involving \(n_w\) Gradient Generation Units (GGUs). At each iteration \(t\), each unit produces a message, and these messages are aggregated using a function \(\text{Agg}\), and the model parameters are then updated using that aggregated vector. Figure~\ref{fig:Setting} illustrates this setup.

This abstraction covers centralized mini-batch training, distributed synchronous training, and simplified federated-learning settings in which client updates are averaged. Surprisingly, although our theoretical derivation is specific to the mean operator, our attack shows good performance against coordinate-wise median and trimmed mean aggregators.

\paragraph{Attacker's write access.}
Within this distributed framework, consider that at every training epoch \(t\), a malicious actor is hidden among the workers and contributes a subset of examples to the final batch. Let \(K_t^H\subset D_t^H\) denote the clean examples received by the compromised worker before label manipulation, and let
$
k_t=\frac{|K_t^H|}{|D_t^H|}
$
denote the adversary's write-access fraction at round \(t\). Unless otherwise stated, we use \(k\) for a fixed write-access rate.

The adversary may replace the clean label \(y_i^H\) of a controlled point \(i\in K_t^H\) by another feasible label \(\tilde y_i\in\mathcal Y\). However, the adversary cannot modify the feature vector \(x_i\), inject new examples, remove honest examples, directly overwrite gradients, or change the server-side optimizer. The corrupted set sent to the learner is
\[
K_t(\tilde y)=\{(x_i,\tilde y_i): i\in K_t^H\},
\qquad
D_t(\tilde y)=H_t\cup K_t(\tilde y).
\]



\paragraph{Local and global budgets.}
The adversary is constrained by a per-round local flipping budget \(b\in[0,1]\):
\[
\sum_{i\in K_t^H}\mathbf 1[\tilde y_i\neq y_i^H] \leq \lfloor b|K_t^H|\rfloor .
\]
Thus, at most a fraction \(b\) of the labels under adversarial write access can be changed at any round. The total corrupted fraction of the full batch is at most
$
\rho = k b ,
$
which we call the global corruption budget. Analyzing the interplay between \(k\) and \(b\) is of interest because for the same product \(kb\), it is important to know which is preferable: a small adversarial region with many flips and a broad adversarial region.



\paragraph{On the weak adversary}

Throughout this work, \textit{weaker adversary} refers strictly to an action-constrained adversary: the attacker may only flip labels (under a strict budget, no feature modification, no arbitrary gradient injection). 
Prior work on gradient-based attacks~\cite{bouaziz2024invertinggradientattacksmakes,baruch2019little} assumes that the adversary is omniscient, therefore, we allow the attacker to be \emph{omniscient}: they have full read-access to the model parameters at every epoch. 
We adopt an omniscient adversary to align with prevalent threat models in gradient-based poisoning/Byzantine-robust literature~\cite{lilisu2018distribstatml, mhamdi2018hidden, DBLP:journals/corr/abs-1802-10116}.
For the more realistic limited-knowledge case, Appendix~\ref{appdx:On_the_threat_model} discusses how an attacker can conceptually train a local surrogate on its accessible subset \(K_t^H\) and apply the same greedy selection there.





\subsection{Label Flipping as a Constrained Optimization Problem}
\label{sec:lfpoison}

Let
$
L_{D_H}(\theta_t)\ = \frac{1}{|D_t^H|}\sum_{i\in D_t^H}\ell(\theta_t;x_i,y_i^H)
$
be the clean empirical risk at round \(t\), and let
$
g_t^H = \nabla L_{D_H}(\theta_t)\
$
be the clean gradient.
Given a label assignment \(\tilde y\) satisfying the budget constraint, define the poisoned gradient
\[
g_t(\tilde y)
=
\frac{1}{|D_t^H|}
\left(
\sum_{i\in H_t}\nabla_\theta \ell(\theta_t;x_i,y_i^H)
+
\sum_{i\in K_t^H}\nabla_\theta \ell(\theta_t;x_i,\tilde y_i)
\right).
\]
At epoch \(t\), the server would normally evaluate the empirical loss on the \textit{honest} batch  
and update the model with the corresponding gradient \(\nabla L_{D_H}(\theta_t)\).  
However, once the man‑in‑the‑middle adversary flips some labels, the server instead observes the \textit{poisoned} gradient $g_t(\tilde y)$
where \(\ell\) is the per‑sample cross‑entropy loss.


The long-horizon availability objective is
$
\max_{\tilde y_0,\ldots,\tilde y_{T-1}}
\mathcal E(\theta_T),
$
where \(\mathcal E\) is a clean validation loss 
and \(\theta_T\) is the model obtained after training under the sequence of poisoned gradients.

This objective is combinatorial and depends on the entire future trajectory.
\aekedit{
We therefore use a tractable one-step surrogate: at each round, the attacker chooses labels so that the poisoned gradient is poorly aligned with a chosen reference direction \(\Delta_t\).
}
Formally, the attacker solves
\begin{figure}[htb]
\vspace{-\baselineskip}
\centering
\begin{minipage}[t]{0.5\textwidth}
\vspace{-4\baselineskip}
\begin{tcolorbox}[colback=iclrblue, colframe=iclrteal, boxrule=1pt, arc=4pt, left=2pt, right=2pt, top=2pt, bottom=2pt]
\begin{subequations}\label{eq:lf_opt}
\begin{align}
\arg\min_{\{\tilde y_i\}_{i\in K_t^H}}
&\quad
\left\langle \Delta_t,\; g_t(\tilde y)\right\rangle
\tag{1}
\label{eq:lf_obj}
\\[-1mm]
\text{s.t.}
&\quad
\underbrace{
\sum_{i\in K_t^H}
\mathbf{1}\!\left[\tilde y_i\neq y_i^H\right]
\le
\lfloor b|K_t^H|\rfloor
}_{\text{local flipping budget}}
\tag{BC}
\label{cons:budget}
\\[-1mm]
&\quad
\tilde y_i\in\mathcal Y,
\qquad
\forall i\in K_t^H .
\notag
\end{align}
\end{subequations}
\end{tcolorbox}
\end{minipage}\hfill
\begin{minipage}[htb]{0.49\textwidth}
\centering
\begin{tikzpicture}[scale=0.65]
    \coordinate (O) at (0,0);
    \coordinate (clean) at (-1.0,1.5);
    \coordinate (poison) at (-0.4,-1.7);
    \coordinate (target) at (1.3,1.6);

    \draw[->, red, thick] (O) -- (clean) node[midway,left] {$-g_t^H$};
    \draw[->, green!60!black, thick] (O) -- (poison) node[midway,right] {$-g_t(\tilde y)$};
    \draw[->, black, thick] (O) -- (target) node[midway,right] {$\theta^{\mathrm{tar}}-\theta_t$};

    \fill (O) circle (1.3pt);
    \node at (-0.15,-0.25) {$\theta_t$};
    \node at (1.55,1.85) {$\theta^{\mathrm{tar}}$};
\end{tikzpicture}
\captionof{figure}{Desired training step direction of honest workers (red), and that of the targeted (black) and untargeted (green) attackers.}
\label{fig:attack_diagram}
\end{minipage}
\end{figure}




The reference direction \(\Delta_t\) specifies the attacker's goal. We distinguish two goals:
\[
\Delta_t
=
\begin{cases}
g_t^H,
&
\text{untargeted attack},
\\[4pt]
\theta^{\text{Target}}-\theta_t,
&
\text{target-parameter attack}. 
\end{cases}
\tag{2}
\label{eq:reference_direction}
\]


The untargeted adversary tries to deviate from the honest gradient,
whereas the targeted adversary steers the update toward a pre‑selected
parameter vector \(\theta^{\text{Target}}\).  Figure~\ref{fig:attack_diagram}
illustrates how the targeted variant uses the vector
\((\theta^{\text{Target}}-\theta_t)\) to bias each gradient step toward
\(\theta^{\text{Target}}\). 

Since the problem is solved independently at each step, we omit the temporal index in the remainder of the paper. \aeedit{\textit{In the rest of our work, we focus on untargeted attacks}} and address Problem~\ref{eq:lf_opt} for multi-class softmax regression.
\section{Multi-Class Softmax Regression}
\label{sec:Multiclass_case}

We now solve Problem~\ref{eq:lf_opt} in the general classification setting. 
Let \(\mathcal Y=\{1,\ldots,C\}\), and let
$
W_t\in\mathbb{R}^{C\times(d+1)}
$
be the current parameter matrix. For each example \(x^{(n)}\), define
\[
p^{(n)}
=
\operatorname{softmax}(W_t x^{(n)})\in\mathbb{R}^C,
\qquad
p^{(n)}_j
=
\operatorname{softmax}(W_t x^{(n)})_j .
\]
We encode labels using a one-hot matrix \(T\in\{0,1\}^{C\times N}\), where
\[
T_{cn}=1
\quad\Longleftrightarrow\quad
\text{example }n\text{ is assigned label }c.
\]
Thus each column of \(T\) satisfies
$
\sum_{c=1}^C T_{cn}=1.
$

The cross-entropy loss on a batch \(D\) can be written as
\[
L_D(W_t;T)
=
-\frac{1}{N}
\sum_{n=1}^{N}
\sum_{c=1}^{C}
T_{cn}\log p^{(n)}_c .
\]
The gradient with respect to the \(j\)-th row of \(W_t\) is
\begin{align*}    
\nabla_{W_j} L_D(W_t;T)
&= - \frac{1}{N} \sum_{n=1}^{N} \sum_{c=1}^{C} T_{cn} \left( \mathds{1}[c=j] - p^{(n)}_j\right) \mathbf{ x}^{(n)} 
\\
&=
\frac{1}{N}
\sum_{n=1}^{N}
\left(
p^{(n)}_j-T_{jn}
\right)
x^{(n)} 
\end{align*}


Let
$
\Delta_t\in\mathbb{R}^{C\times(d+1)}
$
be the reference direction from Problem~\ref{eq:lf_opt}. Recall
$$
\Delta_t
=
\begin{cases}
g_t^H,
&
\text{untargeted attack},
\\[4pt]
W^{\mathrm{tar}}-W_t,
&
\text{targeted attack}.
\end{cases}
$$
The inner product is now the Frobenius inner product:
\[
\left\langle \Delta_t,\;g_t(T)\right\rangle_F
=
\sum_{j=1}^C
\left\langle
\Delta_{t,j},
\nabla_{W_j}L_D(W_t;T)
\right\rangle ,
\]
where \(\Delta_{t,j}\) is the \(j\)-th row of \(\Delta_t\).

Using the expression for the gradient, we obtain
\[
\left\langle \Delta_t,\;g_t(T)\right\rangle_F
=
\frac{1}{N}
\sum_{n=1}^{N}
\sum_{j=1}^{C}
\left(
p^{(n)}_j-T_{jn}
\right)
\left\langle
\Delta_{t,j},x^{(n)}
\right\rangle .
\]
For honest examples \(n\in H_t\), the label is fixed. Only the columns \(n\in K_t^H\) are decision variables for the attacker. Dropping all terms independent of the attacker's choices, the label-dependent objective becomes
\[
-\frac{1}{N}
\sum_{n\in K_t^H}
\sum_{c=1}^{C}
T_{cn}
\left\langle
\Delta_{t,c},x^{(n)}
\right\rangle .
\]
Therefore, for each controlled example \(n\), define the class score
$
Z_{cn}
=
-\left\langle
\Delta_{t,c},x^{(n)}
\right\rangle .
$
Then the multi-class label-flipping problem is equivalent to
\begin{tcolorbox}[colback=iclrblue, colframe=iclrteal, boxrule=1pt, arc=4pt, left=-1pt, right=4pt, top=2pt, bottom=2pt]
\[
\begin{aligned}
\arg\min_{T}
&\quad
\sum_{n\in K_t^H}
\sum_{c=1}^{C}
T_{cn}Z_{cn}
\\
\text{s.t.}
&\quad
\sum_{c=1}^{C}T_{cn}=1,
\qquad
\forall n\in K_t^H,
\\
&\quad
T_{cn}\in\{0,1\},
\qquad
\forall c,n,
\\
&\quad
\sum_{n\in K_t^H}
\left(
1-T_{y_n^H,n}
\right)
\le
\lfloor b|K_t^H|\rfloor .
\end{aligned}
\tag{MC}
\label{eq:multiclass_lf}
\]
\end{tcolorbox}

For each controlled example \(n\), the best label ignoring the budget is
\[
c^\star(n)
\in
\arg\min_{c\in\{1,\ldots,C\}}
Z_{cn}
=
\arg\max_{c\in\{1,\ldots,C\}}
\left\langle
\Delta_{t,c},x^{(n)}
\right\rangle .
\]
Define the improvement over the clean label:
\[
\Gamma_n
=
Z_{y_n^H,n}
-
Z_{c^\star(n),n}.
\]
A flip is beneficial exactly when \(\Gamma_n>0\). Thus, under budget
\[
p=\lfloor b|K_t^H|\rfloor,
\]
the optimal one-step attack changes the labels of the \(p\) controlled examples with the largest positive values of \(\Gamma_n\), assigning each selected example \(n\) to \(c^\star(n)\).

\begin{algorithm}[H]
\caption{Per-round greedy label selection for multi-class softmax regression}
\label{algo:multiclass}
\begin{algorithmic}[1]
\REQUIRE Controlled clean set \(K_t^H=\{(x^{(n)},y_n^H)\}_{n\in K_t^H}\), current weight matrix \(W_t\), reference matrix \(\Delta_t\), budget \(b\)
\ENSURE Poisoned labels \(\{\tilde y_n\}_{n\in K_t^H}\)
\STATE \(p\leftarrow \lfloor b|K_t^H|\rfloor\)
\FOR{each \(n\in K_t^H\)}
    \FOR{each class \(c\in\{1,\ldots,C\}\)}
        \STATE \(Z_{cn}\leftarrow -\langle \Delta_{t,c},x^{(n)}\rangle\)
    \ENDFOR
    \STATE \(c^\star(n)\leftarrow \arg\min_c Z_{cn}\)
    \STATE \(\Gamma_n\leftarrow Z_{y_n^H,n}-Z_{c^\star(n),n}\)
    \STATE Initialize \(\tilde y_n\leftarrow y_n^H\)
\ENDFOR
\STATE Let \(S\) be the indices of the \(p\) largest positive values of \(\Gamma_n\)
\FOR{each \(n\in S\)}
    \STATE \(\tilde y_n\leftarrow c^\star(n)\)
\ENDFOR
\STATE \textbf{return} \(\{\tilde y_n\}_{n\in K_t^H}\)
\end{algorithmic}
\end{algorithm}

\begin{theorem}[\aekedit{Exact optimality for the per-round directional surrogate objective under mean aggregation in the multi-class setting}]
For multi-class softmax regression under mean aggregation, Algorithm~\ref{algo:multiclass} solves Problem~\ref{eq:lf_opt} exactly at round \(t\) for any fixed reference direction \(\Delta_t\).
\end{theorem}

\begin{proof}
At round \(t\), the objective
\[
\left\langle \Delta_t,g_t(T)\right\rangle_F
\]
can be written as a constant independent of the attack plus
\[
\frac{1}{N}
\sum_{n\in K_t^H}
\sum_{c=1}^{C}
T_{cn}Z_{cn}.
\]
Thus, for each controlled example \(n\), the best label ignoring the budget is
\[
c^\star(n)\in\arg\min_c Z_{cn}.
\]
Changing the clean label \(y_n^H\) to this best label reduces the objective by
\[
\Gamma_n
=
Z_{y_n^H,n}-Z_{c^\star(n),n}.
\]
The only coupling between examples is the cardinality budget
\[
\sum_{n\in K_t^H}
(1-T_{y_n^H,n})
\le
\lfloor b|K_t^H|\rfloor .
\]
Therefore, the optimal feasible solution is obtained by selecting the largest positive improvements \(\Gamma_n\), up to the budget. This is exactly Algorithm~\ref{algo:multiclass}.
\end{proof}

This result shows that the multi-class attack is not a separate construction: it is the direct softmax-regression instance of the same directional label-flipping problem introduced in Section~\ref{sec:lfpoison}.

\newcommand{\plotroot}{full_paper_sections/comparison_with_random}

\section{Multi-Class Experimental Results}
\label{sec:multiclass_results}

\paragraph{Experimental setup.}
We evaluate the multi-class attack on MNIST10 and CIFAR10.
\aeedit{Since unit-norm feature normalization is commonly used when feature vectors are fed to linear classifiers, nearest-neighbor classifiers, or dot-product-based losses (~\cite{Tian2020RethinkingFI,pmlr-v139-radford21a,khosla2021supervisedcontrastivelearning,pmlr-v119-chen20j}), we evaluate the attack on normalized inputs.}
Unless stated otherwise, the main experiments use multi-class logistic regression trained with SGD under mean aggregation, matching the setting analyzed in Section~\ref{sec:Multiclass_case}. We also report results for a two-layer MLP to test whether the same label-selection rule remains effective beyond the convex softmax-regression setting. Data are split iid across workers, and all results are averaged over six independent runs. 

We compare our attack with random label flipping under the same \(k\), \(b\), dataset, model, and training protocol. In the random baseline, the attacker flips the same number of labels as the optimized attack but selects examples and replacement labels without using the gradient-obstruction score. By this, we test whether degradation comes from the optimized choice of labels or from injecting label noise.

\begin{figure*}[htb]
    \centering

    \begin{subfigure}[htb]{0.48\textwidth}
        \centering
        \includegraphics[width=\linewidth]{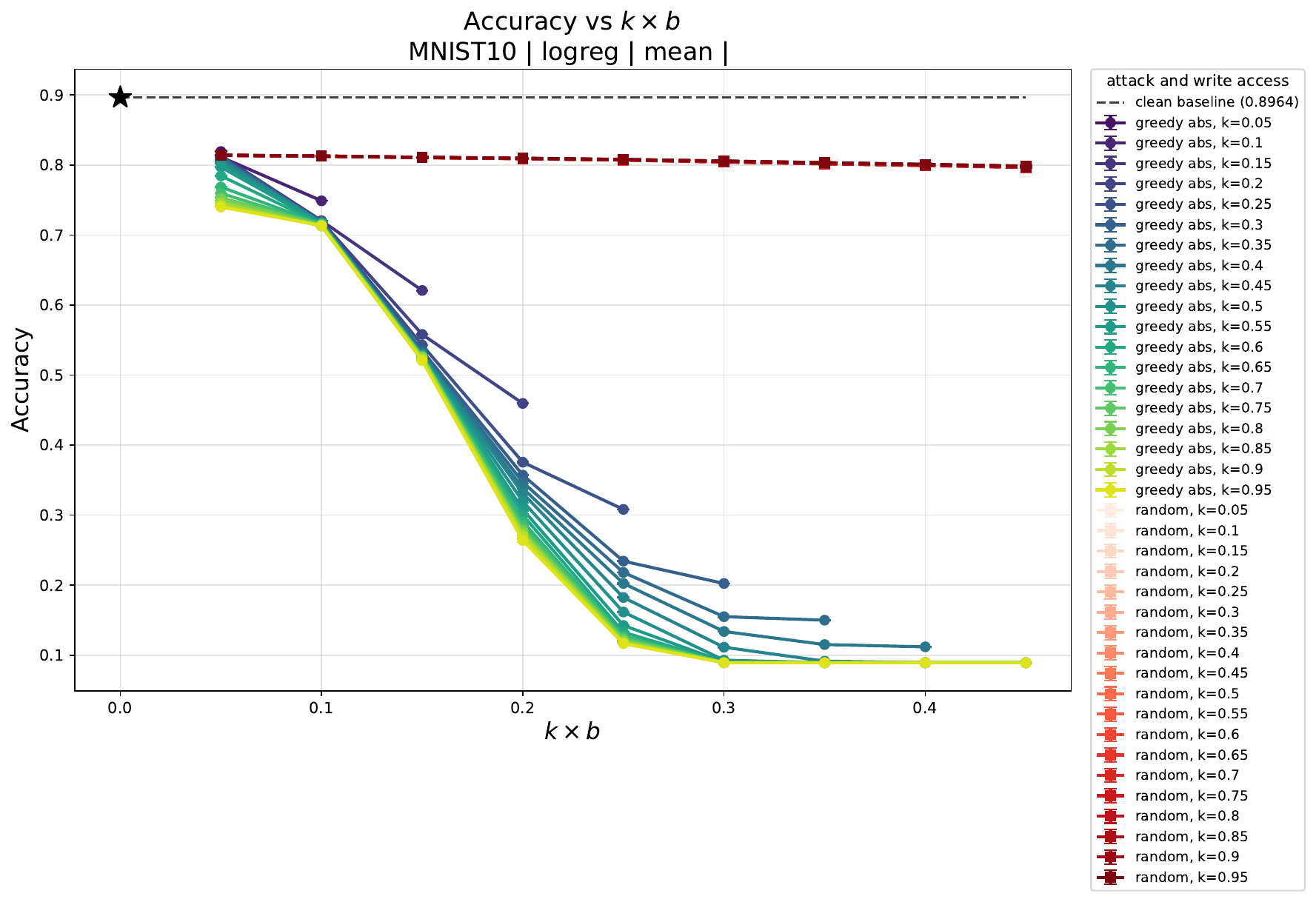}
        \caption{MNIST10, logistic regression.}
        \label{fig:norm-mnist10-logreg-kb}
    \end{subfigure}
    \hfill
    \begin{subfigure}[htb]{0.48\textwidth}
        \centering
        \includegraphics[width=\linewidth]{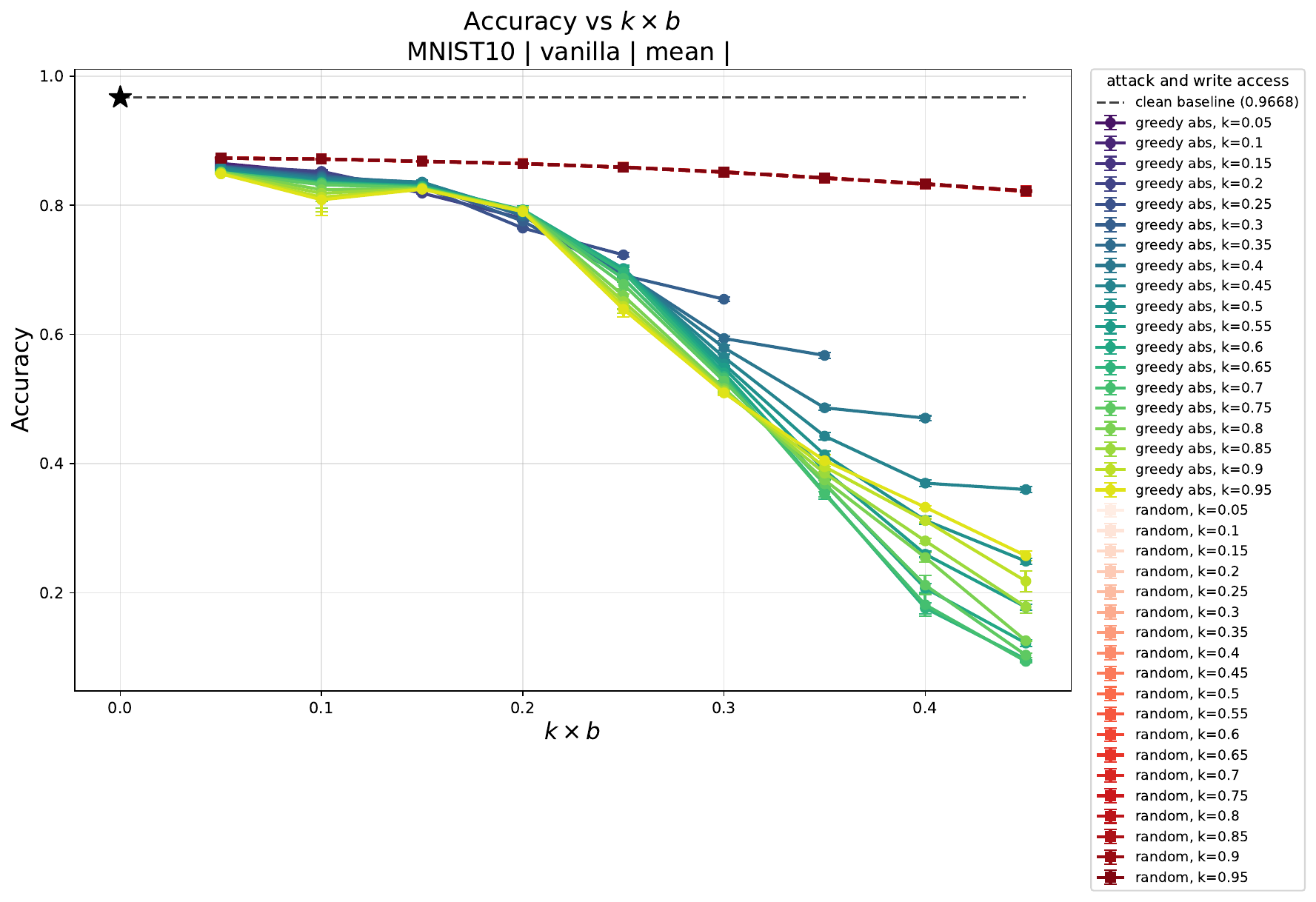}
        \caption{MNIST10, two-layer MLP.}
        \label{fig:norm-mnist10-mlp-kb}
    \end{subfigure}

    \vspace{0.75em}

    \begin{subfigure}[htb]{0.48\textwidth}
        \centering
        \includegraphics[width=\linewidth]{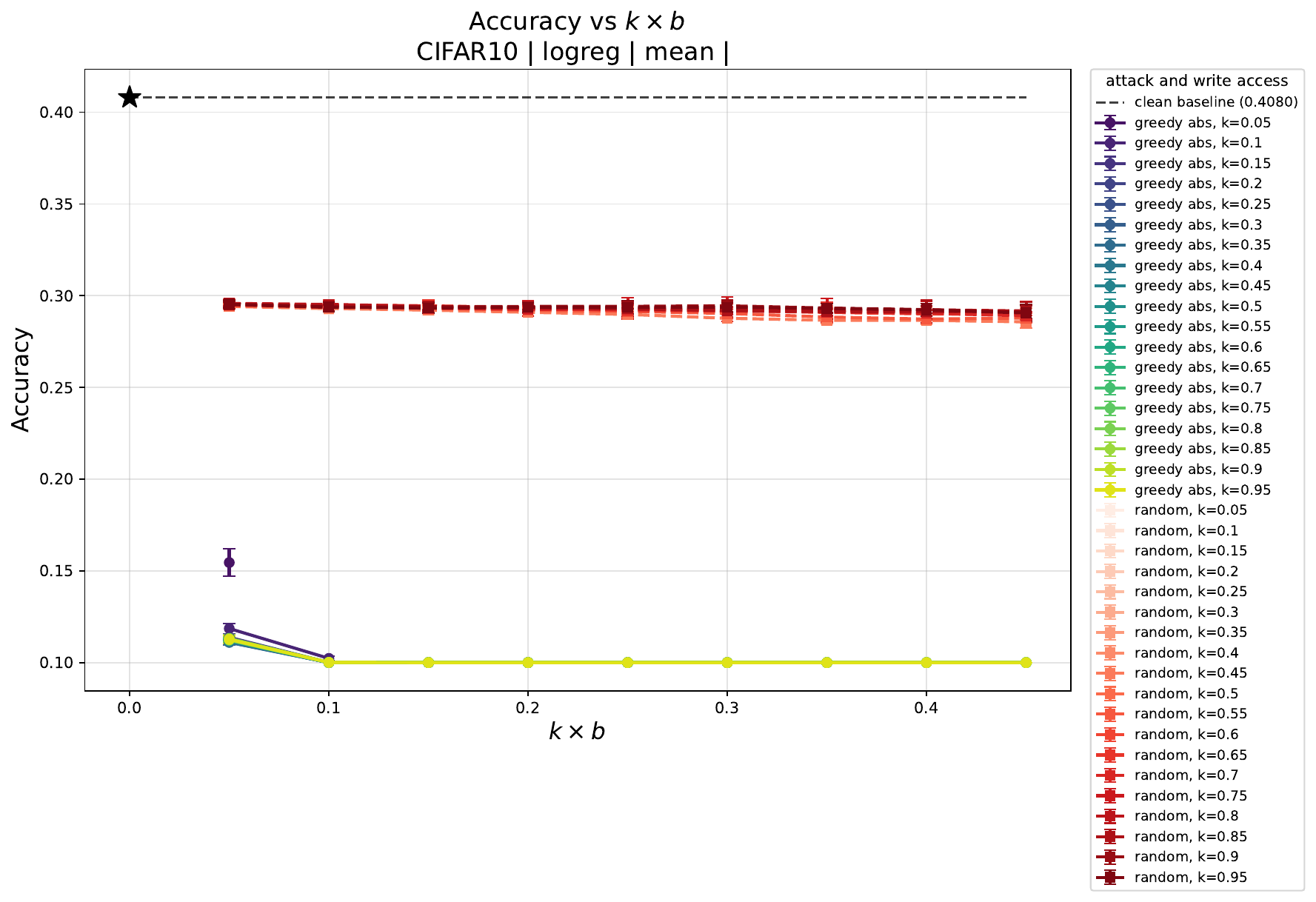}
        \caption{CIFAR10, logistic regression.}
        \label{fig:norm-cifar10-logreg-kb}
    \end{subfigure}
    \hfill
    \begin{subfigure}[htb]{0.48\textwidth}
        \centering
        \includegraphics[width=\linewidth]{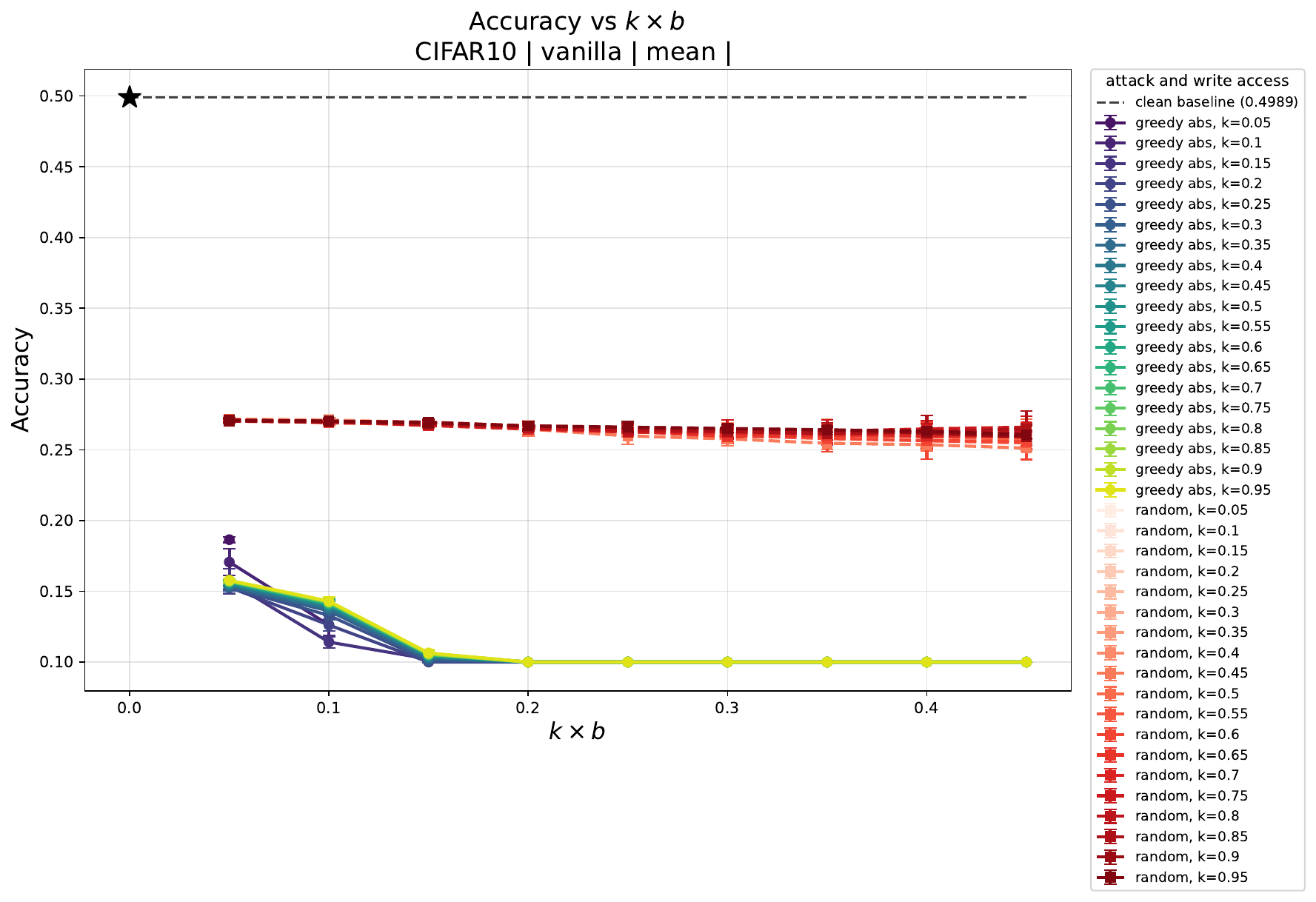}
        \caption{CIFAR10, two-layer MLP.}
        \label{fig:norm-cifar10-mlp-kb}
    \end{subfigure}

    \caption{
    Accuracy versus effective attack budget \(k\times b\) for normalized inputs. 
    Our attack induces stronger degradation than random flipping, showing that the gradient-obstruction objective provides information beyond random label noise.
    }
    \label{fig:normalized-accuracy-vs-kb}
\end{figure*}


\paragraph{Optimized flipping causes stronger availability degradation than random noise.}
Figure~\ref{fig:normalized-accuracy-vs-kb} reports test accuracy as a function of the effective corruption budget $(\rho = k b)$. Across the tested multi-class settings, the optimized attack generally reduces accuracy more than random flipping at the same budget. This shows that the greedy score identifies labels whose changes are harmful to the training direction, rather than merely injecting noise.

The effect is visible at small global budgets. For example, in all tested datasets, changing only a small fraction of the full batch produces a measurable drop in clean test accuracy. As the budget increases, the gap between the clean baseline and the poisoned runs widens, which is natural since larger budgets give the attacker more power.

Notably, although the attack was tailored to logistic regression, it works well on a multi-layer perceptron under the setting described in section \ref{how-deeper-models}.


\subsection{Effect of Write Access and Local Budget}

The parameters \(k\) and \(b\) capture different aspects of the adversary's power. The write-access fraction \(k\) determines how many examples the attacker can inspect and potentially modify, while the local budget \(b\) determines how many of those controlled examples may actually be flipped. Although the product \(k b\) gives the global corruption rate, the attack is not determined only by this product.

\begin{wrapfigure}{r}{0.4\textwidth}
        \centering
      \vspace{-2\baselineskip}
        \includegraphics[width=0.5\textwidth]{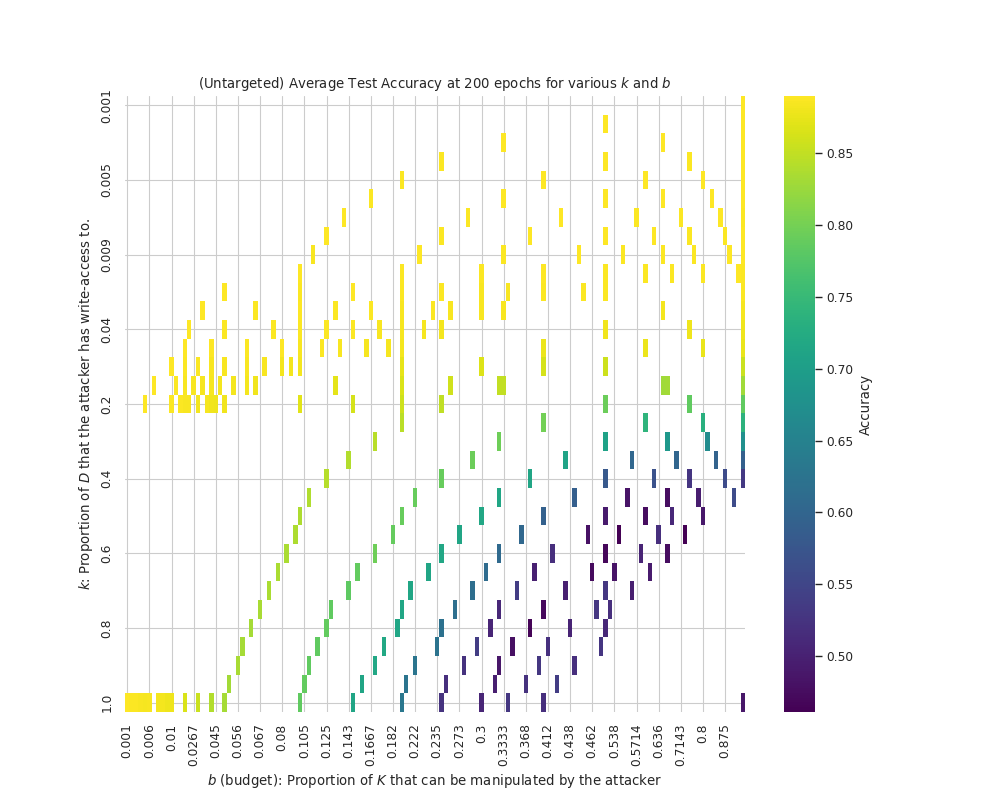}
        \caption{Heatmap of test accuracy as a function of $k$ and $b$ in a binary classification setting  between 2 MNIST classes.}
        \label{fig:heatmap_untargeted}
  \vspace{-\baselineskip}
\end{wrapfigure}

To visualize this effect more directly, Figure~\ref{fig:heatmap_untargeted} reports a heatmap of the accuracy as a function of $k$ and $b$ on a binary MNIST (classes 0 and 1) dataset. Although this setting is simpler than the multi-class experiments above, it isolates the roles of \(k\) and \(b\).

\label{sec:k_b_tradeoff}




Empirically, Figures~\ref{fig:normalized-accuracy-vs-kb} and~\ref{fig:heatmap_untargeted} show that, for a fixed global corruption rate $\rho = kb$, increasing the write-access fraction $k$ is often more damaging than increasing the local flipping budget $b$. In particular, when $k$ is small, increasing $b$ has limited effect on test accuracy, whereas larger values of $k$ consistently make the attack more effective.
This suggests that, for a fixed total flipping proportion $\rho$, broad write access is more valuable to the attacker than the ability to flip many labels within a narrow controlled subset.

This behavior can be explained by the structure of the per-round optimization problem. A larger controlled set $K_t^H$ gives the attacker a richer pool of candidate examples and therefore increases the probability of finding labels with large anti-alignment scores. By contrast, increasing $b$ while keeping $k$ fixed only allows the attacker to flip more labels within the same candidate pool, which may already contain only a limited number of highly damaging examples. 
This can also be understood through the attainable gradient set. Since the gradient is a weighted combination of feature vectors and the label-dependent weights are discrete, a larger write-access set expands the finite set of poisoned gradient directions available to the attacker.






\subsection{Transfer to Robust Aggregation}
\label{sec:robust_aggregators}






\begin{figure*}[htb]
    \centering

    \begin{subfigure}[htb]{0.48\textwidth}
        \centering
        \includegraphics[width=\linewidth]{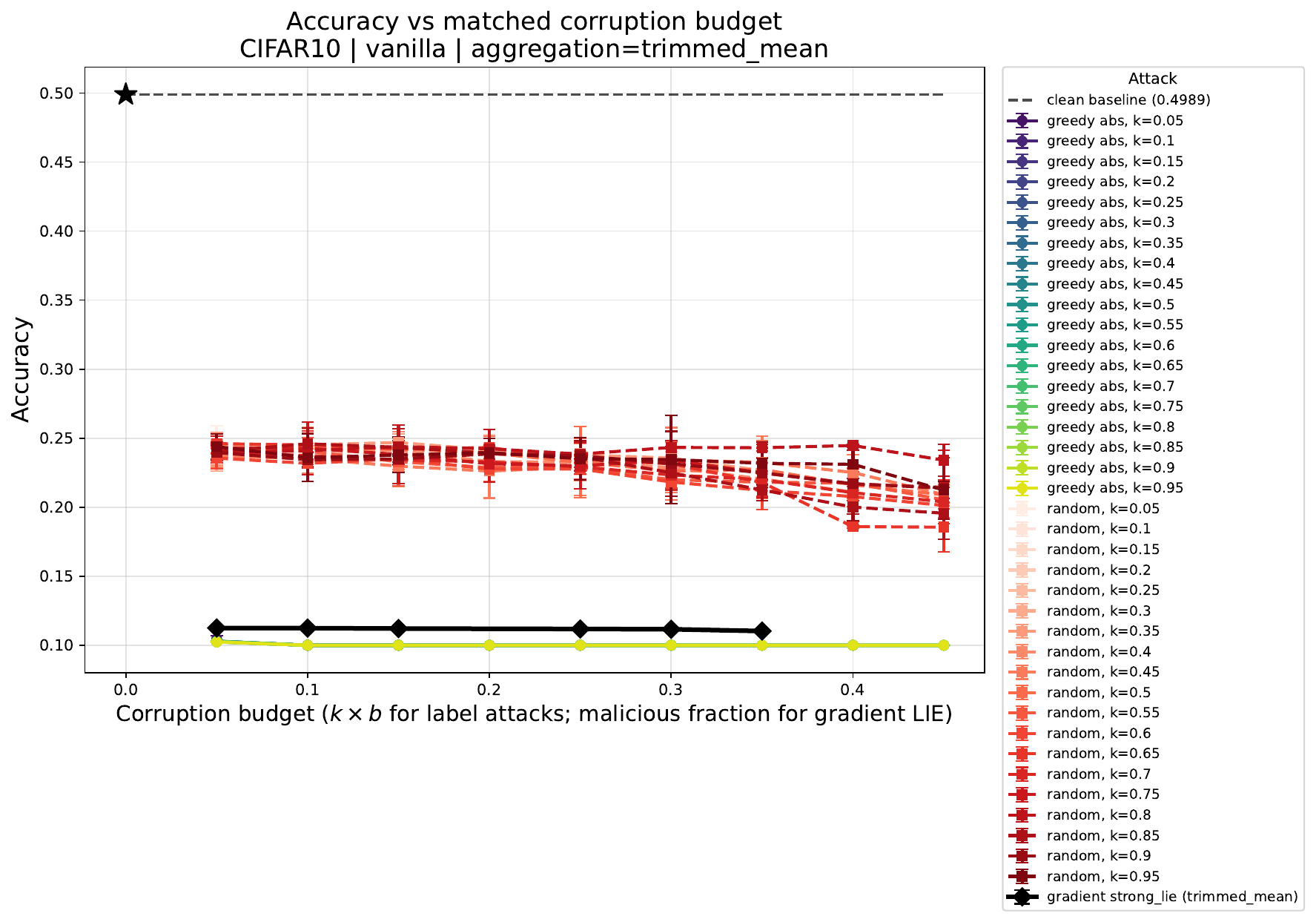}
        \caption{CIFAR10, two-layer MLP, trimmed mean aggregator.}
    \end{subfigure}
    \hfill
    \begin{subfigure}[htb]{0.48\textwidth}
        \centering
        \includegraphics[width=\linewidth]{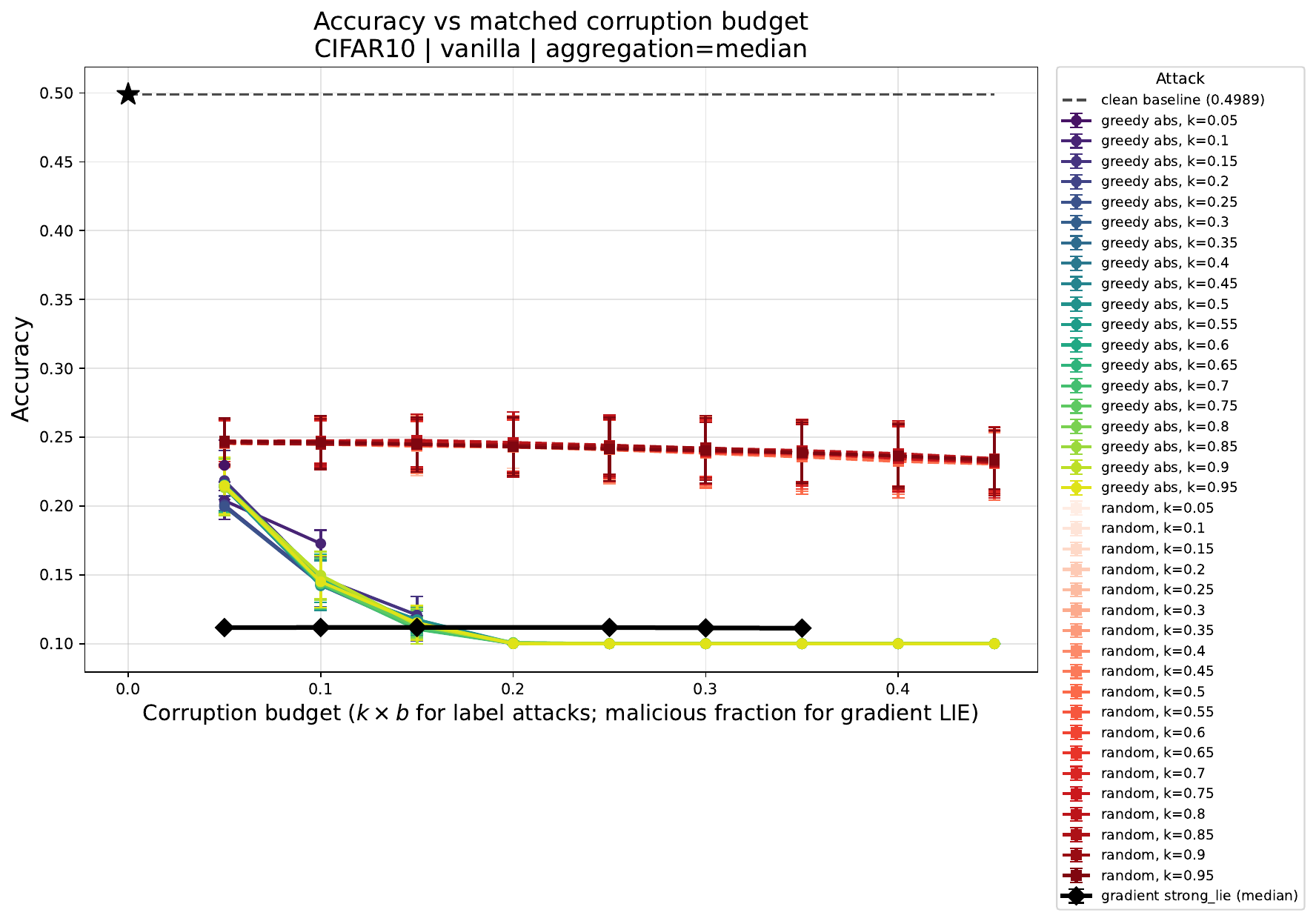}
        \caption{CIFAR10, two-layer MLP, coordinate-wise median aggregator.}
    \end{subfigure}

    \caption{
    Test accuracy under trimmed mean (left), and coordinate-wise median (right) on normalized CIFAR10 with a two-layer MLP. Although the attack is derived for mean aggregation, optimized label flipping degrades accuracy under these robust aggregators, and approaches the degradation of the Little-is-Enough gradient attack in the tested setting.
    }
    \label{fig:robust_agg_cifar}
\end{figure*}
Our theoretical derivation and one-step optimality result are specific to mean aggregation. Nevertheless, we also test whether the same label-selection rule remains effective when the server replaces the mean with a coordinate-wise median or trimmed-mean aggregators. 

\begin{figure*}[htb]
    \centering

    \begin{subfigure}[htb]{0.48\textwidth}
        \centering
        \includegraphics[width=\linewidth]{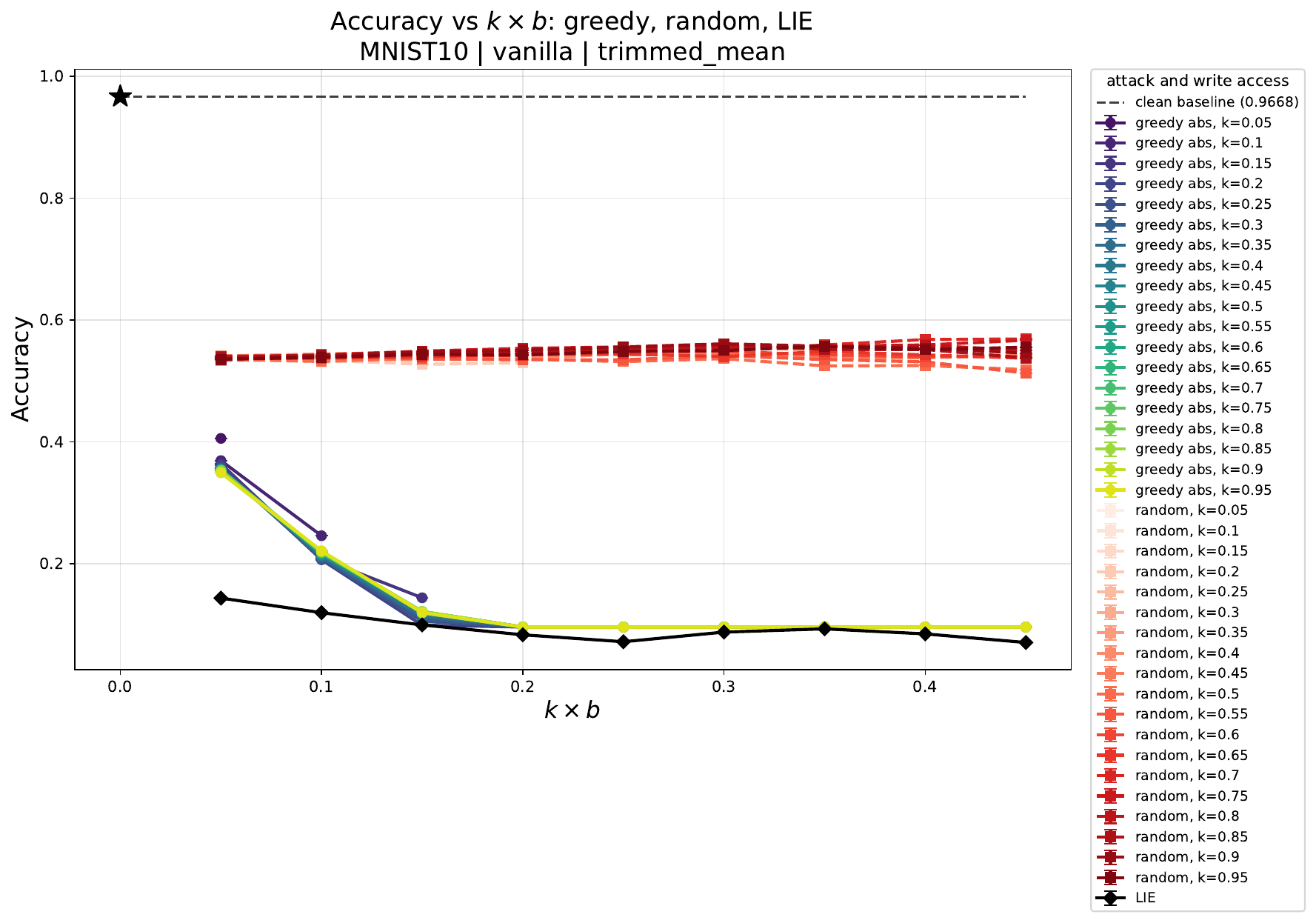}
        \caption{\aekedit{MNIST10, two-layer MLP, trimmed mean aggregator.}}
    \end{subfigure}
    \hfill
    \begin{subfigure}[htb]{0.48\textwidth}
        \centering
        \includegraphics[width=\linewidth]{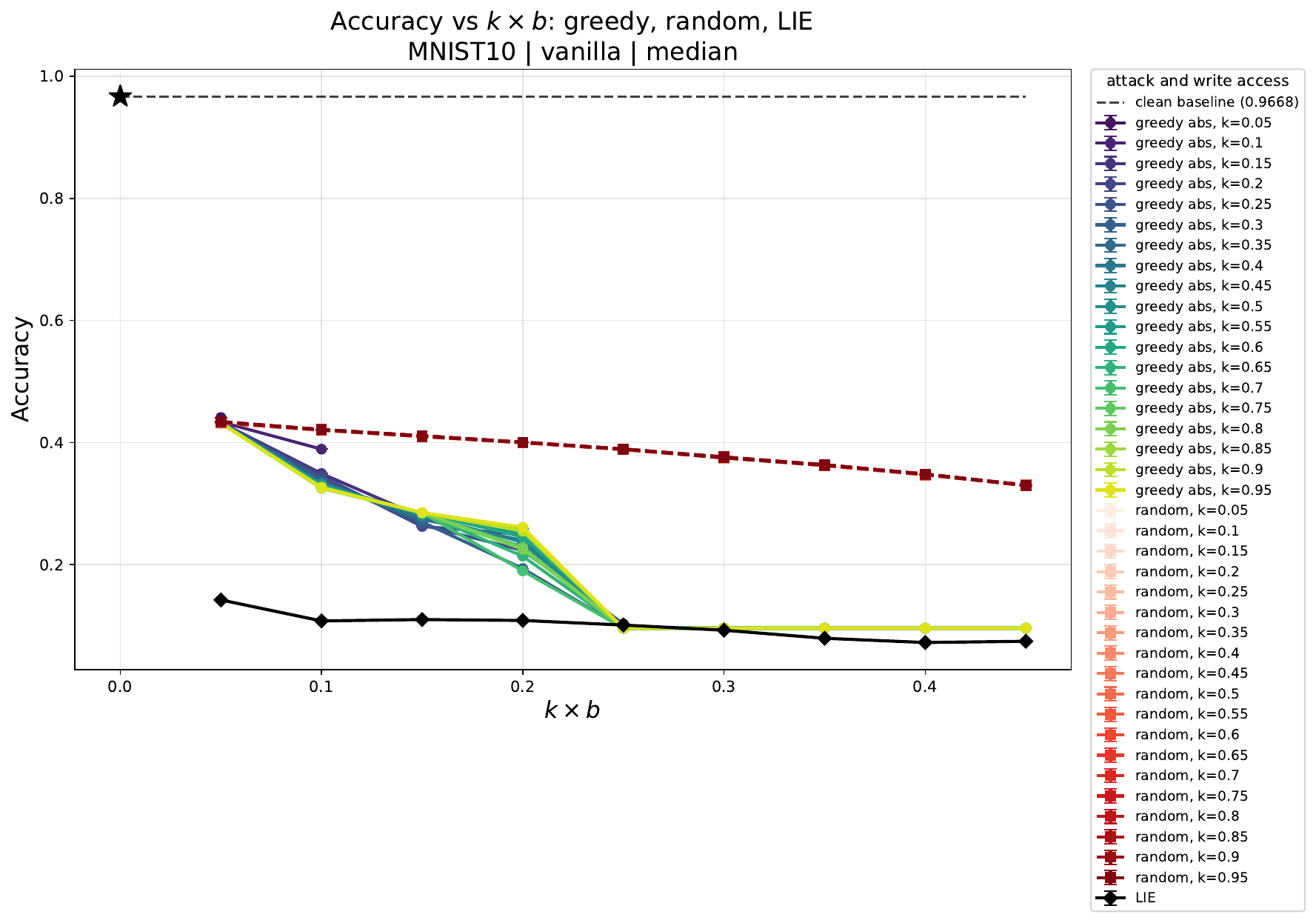}
        \caption{\aekedit{MNIST10, two-layer MLP, coordinate-wise median aggregator.}}
    \end{subfigure}

    \caption{
    \aekedit{Test accuracy under trimmed mean (left), and coordinate-wise median (right) on normalized MNIST10 with a two-layer MLP. Although the attack is derived for mean aggregation, optimized label flipping degrades accuracy under these robust aggregators, and approaches the degradation of the Little-is-Enough gradient attack in the tested setting.}
    }
    \label{fig:robust_agg_mnist}

\end{figure*}

\aeedit{Surprisingly}, Figures~\ref{fig:robust_agg_cifar} and~\ref{fig:robust_agg_mnist} shows that the proposed label-flipping strategy remains \aeedit{damaging} under both robust aggregation rules, even though it was only designed to attack the non-robust mean aggregation. This can be explained by the restricted nature of the attack. Unlike arbitrary gradient attacks, the adversary does not submit an unconstrained update vector that can be easily identified as an outlier. Instead, each malicious update is induced by valid training examples whose features are unchanged and whose labels remain in the feasible label set. As a result, the poisoned gradients can remain close enough to the distribution of honest gradients that the tested robust aggregators do not fully remove their effect.

In contrast, random label flipping is consistently weaker, showing that the degradation is not merely caused by adding label noise. We also include the Little-is-Enough gradient attack for comparison. 
\aekedit{
In Figures~\ref{fig:robust_agg_cifar} and~\ref{fig:robust_agg_mnist}, the horizontal axis's meaning depends on the attack.
For the label-flipping attacks, the plotted value is $\rho = kb$, where $k$ is the attacker’s write-access fraction and $b$ is the local flipping budget; thus $\rho$ is the maximum fraction of the full batch whose labels are modified at a given round. For the Little-is-Enough gradient attack, the plotted value is the malicious-worker fraction $m/n_w$, where m is the number of malicious workers and $n_w$ is the total number of workers. In our comparison, for each label-flipping budget $\rho$, we set $m \approx \rho n_w$.
The comparison is intended to assess how much degradation a label-only write adversary can approach under the same corrupted fraction.

}
Under this interpretation, the results indicate that a label-only adversary can, in these settings, approach the degradation caused by a substantially more powerful gradient-manipulation adversary.

\section{Conclusion}\label{sec:conclusion}

We studied availability attacks under a highly constrained label-flipping threat model. By introducing an intuitive, budget-aware objective, we reveal a vulnerability often studied under stronger adversarial capabilities. Unlike gradient attacks or more general data-poisoning attacks, the adversary considered here cannot modify features, inject samples, delete data, or directly overwrite gradients; it may only flip a limited number of labels among examples under its write access.
Our experiments show that optimized label flipping can substantially degrade test accuracy and consistently outperforms random label flipping under matched budgets. These findings establish a foundation for stronger defenses and, more broadly, a deeper understanding of security in federated and distributed learning. 
Finally, although our theoretical derivation is specific to mean aggregation, the attack transfers empirically to neural networks and robust aggregators such as coordinate-wise median and trimmed mean.
Interesting follow-up directions include
(i) searching for globally optimal flipping attacks, and
(ii) devising practical defenses tailored to the proposed threat model.  

\section*{Broader Impact Statement}

In this paper, we improve the community's understanding of the capabilities of label flipping in damaging the training of a machine learning model. While doing so requires the study and the development of train-phase adversarial attacks, we believe that these attacks serve a mostly pedagogical role and help alert on the potential damage done by an omniscient attacker, previously thought to be able to do harm only when allowed to send gradients in a distributed/federated learning setup.

\bibliography{main}      

@article{capitaine2024unravelling,
  title={Unravelling in collaborative learning},
  author={Capitaine, Aymeric and Boursier, Etienne and Scheid, Antoine and Moulines, Eric and Jordan, Michael and El-Mhamdi, El-Mahdi and Durmus, Alain},
  journal={Advances in Neural Information Processing Systems},
  volume={37},
  pages={97231--97260},
  year={2024}
}

@inproceedings{ananthakrishnan2024delegating,
  title={Delegating data collection in decentralized machine learning},
  author={Ananthakrishnan, Nivasini and Bates, Stephen and Jordan, Michael and Haghtalab, Nika},
  booktitle={International Conference on Artificial Intelligence and Statistics},
  pages={478--486},
  year={2024},
  organization={PMLR}
}

@article{liu2023transferable,
  title={Transferable availability poisoning attacks},
  author={Liu, Yiyong and Backes, Michael and Zhang, Xiao},
  journal={arXiv:2310.05141},
  year={2023}
}

@inproceedings{Bal2025,
 author = {Bal, Melis Ilayda and Cevher, Volkan and Muehlebach, Michael},
 booktitle = {International Conference on Representation Learning},
 editor = {Y. Yue and A. Garg and N. Peng and F. Sha and R. Yu},
 pages = {16235--16269},
 title = {Adversarial Training for Defense Against Label Poisoning Attacks},
 url = {https://proceedings.iclr.cc/paper_files/paper/2025/file/298c3e32d7d402189444be2ff5d19979-Paper-Conference.pdf},
 volume = {2025},
 year = {2025}
}

@article{
bouaziz2024invertinggradientattacksmakes,
title={Inverting Gradient Attacks Makes Powerful Data Poisoning},
author={Wassim Bouaziz and Nicolas Usunier and El-Mahdi El-Mhamdi},
journal={Transactions on Machine Learning Research},
issn={2835-8856},
year={2025},
url={https://openreview.net/forum?id=Lvy5MjyTh3},
note={}
}

@inproceedings{bouaziz2026winter,
  title={Winter soldier: Backdooring language models at pre-training with indirect data poisoning},
  author={Bouaziz, Wassim and Videau, Mathurin and Usunier, Nicolas and El-Mhamdi, El-Mahdi},
  booktitle={International Conference on Learning Representations},
  volume={2026},
  pages={43695--43713},
  year={2026}
}

@inproceedings{farhadkhani2022equivalence,
    title={An equivalence between data poisoning and byzantine gradient attacks},
    author={Farhadkhani, Sadegh and Hoang, Lê-Nguyên and Villemaud, Oscar},
    booktitle={International Conference on Machine Learning},
    year={2022},
}

@article{baruch2019little,
    title={A little is enough: Circumventing defenses for distributed learning},
    author={Baruch, Gilad and Baruch, Moran and Goldberg, Yoav},
    journal={Advances in Neural Information Processing Systems},
    year={2019}
}

@article{blanchard2017byzantinetolerant,
  title={Machine learning with adversaries: Byzantine tolerant gradient descent},
  author={Blanchard, Peva and El Mhamdi, El Mahdi and Guerraoui, Rachid and Stainer, Julien},
  journal={Advances in neural information processing systems},
  volume={30},
  year={2017}
}

@inproceedings{yin2021byzantinerobust,
  title={Byzantine-robust distributed learning: Towards optimal statistical rates},
  author={Yin, Dong and Chen, Yudong and Kannan, Ramchandran and Bartlett, Peter},
  booktitle={International Conference on Machine Learning},
  pages={5650--5659},
  year={2018},
  organization={PMLR}
}

@article{mhamdi2018hidden,
    title={The hidden vulnerability of distributed learning in byzantium},
    author={El-Mhamdi, El-Mahdi and Guerraoui, Rachid and Rouault, S{\'e}bastien},
    journal={International Conference on Machine Learning},
    year={2018},
}

@inproceedings{el2021distributed,
  title={Distributed momentum for byzantine-resilient stochastic gradient descent},
  author={El-Mhamdi, El-Mahdi and Guerraoui, Rachid and Rouault, S{\'e}bastien Louis Alexandre},
  booktitle={9th International Conference on Learning Representations (ICLR)},
  year={2021}
}

@article{geiping2021witches,
  title={Witches' brew: Industrial scale data poisoning via gradient matching},
  author={Geiping, Jonas and Fowl, Liam and Huang, W Ronny and Czaja, Wojciech and Taylor, Gavin and Moeller, Michael and Goldstein, Tom},
  journal={arXiv preprint arXiv:2009.02276},
  year={2020}
}

@article{huang2020metapoison,
  title={Metapoison: Practical general-purpose clean-label data poisoning},
  author={Huang, W Ronny and Geiping, Jonas and Fowl, Liam and Taylor, Gavin and Goldstein, Tom},
  journal={Advances in Neural Information Processing Systems},
  year={2020}
}

@article{steinhardt2017certified,
    title={Certified defenses for data poisoning attacks},
    author={Steinhardt, Jacob and Koh, Pang Wei W and Liang, Percy S},
    journal={Advances in neural information processing systems},
    volume={30},
    year={2017}
}

@article{koh_stronger_2021,
    title={Stronger data poisoning attacks break data sanitization defenses},
    author={Koh, Pang Wei and Steinhardt, Jacob and Liang, Percy},
    journal={Machine Learning},
    pages={1--47},
    year={2022},
    publisher={Springer}
}

@inproceedings{zhao_clpa_2022,
  title = {{{CLPA}}: {{Clean-Label Poisoning Availability Attacks Using Generative Adversarial Nets}}},
  shorttitle = {{{CLPA}}},
  booktitle = {Proceedings of the {{AAAI Conference}} on {{Artificial Intelligence}}},
  author = {Zhao, Bingyin and Lao, Yingjie},
  date = {2022-06-28},
  year = {2022},
  volume = {36},
  pages = {9162--9170},
  issn = {2374-3468, 2159-5399},
  doi = {10.1609/aaai.v36i8.20902},
  url = {https://ojs.aaai.org/index.php/AAAI/article/view/20902},
  urldate = {2023-05-04}
}

@inproceedings{ning2021invisible,
  title={Invisible poison: A blackbox clean label backdoor attack to deep neural networks},
  author={Ning, Rui and Li, Jiang and Xin, Chunsheng and Wu, Hongyi},
  booktitle={IEEE INFOCOM 2021-IEEE Conference on Computer Communications},
  pages={1--10},
  year={2021},
  organization={IEEE}
}

@inproceedings{LablePoisoningIsAllYouNeed,
author = {Jha, Rishi D. and Hayase, Jonathan and Oh, Sewoong},
title = {Label poisoning is all you need},
year = {2023},
publisher = {Curran Associates Inc.},
address = {Red Hook, NY, USA},
booktitle = {Proceedings of the 37th International Conference on Neural Information Processing Systems},
articleno = {3111},
numpages = {24},
location = {New Orleans, LA, USA},
series = {NIPS '23}
}

@article{lilisu2018distribstatml,
    title={Distributed statistical machine learning in adversarial settings: Byzantine gradient descent},
    author={Chen, Yudong and Su, Lili and Xu, Jiaming},
    journal={Proceedings of the ACM on Measurement and Analysis of Computing Systems},
    volume={1},
    number={2},
    pages={1--25},
    year={2017},
    publisher={ACM New York, NY, USA}
}

@article{Lu2022IndiscriminateDP,
  title={Indiscriminate Data Poisoning Attacks on Neural Networks},
  author={Yiwei Lu and Gautam Kamath and Yaoliang Yu},
  journal={ArXiv},
  year={2022},
  volume={abs/2204.09092},
  url={https://api.semanticscholar.org/CorpusID:248266720}
}

@misc{paudice2018labelsanitizationlabelflipping,
      title={Label Sanitization against Label Flipping Poisoning Attacks}, 
      author={Andrea Paudice and Luis Muñoz-González and Emil C. Lupu},
      year={2018},
      eprint={1803.00992},
      archivePrefix={arXiv},
      primaryClass={stat.ML},
      url={https://arxiv.org/abs/1803.00992}, 
}

@article{10.1145/3006384,
author = {Awasthi, Pranjal and Balcan, Maria Florina and Long, Philip M.},
title = {The Power of Localization for Efficiently Learning Linear Separators with Noise},
year = {2017},
issue_date = {February 2017},
publisher = {Association for Computing Machinery},
address = {New York, NY, USA},
volume = {63},
number = {6},
issn = {0004-5411},
url = {https://doi.org/10.1145/3006384},
doi = {10.1145/3006384},
articleno = {50},
numpages = {27}
}

@misc{zhang2017understandingdeeplearningrequires,
      title={Understanding deep learning requires rethinking generalization}, 
      author={Chiyuan Zhang and Samy Bengio and Moritz Hardt and Benjamin Recht and Oriol Vinyals},
      year={2017},
      eprint={1611.03530},
      archivePrefix={arXiv},
      primaryClass={cs.LG},
      url={https://arxiv.org/abs/1611.03530}, 
}

@inproceedings{10.1145/2046684.2046692,
author = {Huang, Ling and Joseph, Anthony D. and Nelson, Blaine and Rubinstein, Benjamin I.P. and Tygar, J. D.},
title = {Adversarial machine learning},
year = {2011},
isbn = {9781450310031},
publisher = {Association for Computing Machinery},
address = {New York, NY, USA},
url = {https://doi.org/10.1145/2046684.2046692},
doi = {10.1145/2046684.2046692},
booktitle = {Proceedings of the 4th ACM Workshop on Security and Artificial Intelligence},
pages = {43–58},
numpages = {16},
location = {Chicago, Illinois, USA},
series = {AISec '11}
}

@misc{shafahi2018poisonfrogstargetedcleanlabel,
      title={Poison Frogs! Targeted Clean-Label Poisoning Attacks on Neural Networks}, 
      author={Ali Shafahi and W. Ronny Huang and Mahyar Najibi and Octavian Suciu and Christoph Studer and Tudor Dumitras and Tom Goldstein},
      year={2018},
      eprint={1804.00792},
      archivePrefix={arXiv},
      primaryClass={cs.LG},
      url={https://arxiv.org/abs/1804.00792}, 
}

@misc{sun2019reallybackdoorfederatedlearning,
      title={Can You Really Backdoor Federated Learning?}, 
      author={Ziteng Sun and Peter Kairouz and Ananda Theertha Suresh and H. Brendan McMahan},
      year={2019},
      eprint={1911.07963},
      archivePrefix={arXiv},
      primaryClass={cs.LG},
      url={https://arxiv.org/abs/1911.07963}, 
}

@inproceedings{DBLP:conf/aistats/McMahanMRHA17,
  author       = {Brendan McMahan and
                  Eider Moore and
                  Daniel Ramage and
                  Seth Hampson and
                  Blaise Ag{\"{u}}era y Arcas},
  editor       = {Aarti Singh and
                  Xiaojin (Jerry) Zhu},
  title        = {Communication-Efficient Learning of Deep Networks from Decentralized
                  Data},
  booktitle    = {Proceedings of the 20th International Conference on Artificial Intelligence
                  and Statistics, {AISTATS} 2017, 20-22 April 2017, Fort Lauderdale,
                  FL, {USA}},
  series       = {Proceedings of Machine Learning Research},
  volume       = {54},
  pages        = {1273--1282},
  publisher    = {{PMLR}},
  year         = {2017},
  url          = {http://proceedings.mlr.press/v54/mcmahan17a.html},
  bibsource    = {dblp computer science bibliography, https://dblp.org}
}

@article{DBLP:journals/corr/Back_to_the_Drawing_Board,
  author       = {Virat Shejwalkar and
                  Amir Houmansadr and
                  Peter Kairouz and
                  Daniel Ramage},
  title        = {Back to the Drawing Board: {A} Critical Evaluation of Poisoning Attacks
                  on Federated Learning},
  journal      = {CoRR},
  volume       = {abs/2108.10241},
  year         = {2021},
  url          = {https://arxiv.org/abs/2108.10241},
  eprinttype    = {arXiv},
  eprint       = {2108.10241},
  bibsource    = {dblp computer science bibliography, https://dblp.org}
}

@inproceedings{Li_quinguru,
author = {Li, Qingru and Wang, Xinru and Wang, Fangwei and Wang, Changguang},
title = {A Label Flipping Attack on Machine Learning Model and Its Defense Mechanism},
year = {2022},
isbn = {978-3-031-22676-2},
publisher = {Springer-Verlag},
address = {Berlin, Heidelberg},
url = {https://doi.org/10.1007/978-3-031-22677-9_26},
doi = {10.1007/978-3-031-22677-9_26},
booktitle = {Algorithms and Architectures for Parallel Processing: 22nd International Conference, ICA3PP 2022, Copenhagen, Denmark, October 10–12, 2022, Proceedings},
pages = {490–506},
numpages = {17},
location = {Copenhagen, Denmark}
}

@inproceedings{Lavaur_systematic,
author = {Lavaur, L\'{e}o and Busnel, Yann and Autrel, Fabien},
title = {Systematic Analysis of Label-flipping Attacks against Federated Learning in Collaborative Intrusion Detection Systems},
year = {2024},
isbn = {9798400717185},
publisher = {Association for Computing Machinery},
address = {New York, NY, USA},
url = {https://doi.org/10.1145/3664476.3670434},
doi = {10.1145/3664476.3670434},
booktitle = {Proceedings of the 19th International Conference on Availability, Reliability and Security},
articleno = {63},
numpages = {12},
location = {Vienna, Austria},
series = {ARES '24}
}

@ARTICLE{GNN_LF,
  author={Yu, Shanqing and Shen, Jie and Xu, Shaocong and Wang, Jinhuan and Wang, Zeyu and Xuan, Qi},
  journal={IEEE Transactions on Network Science and Engineering}, 
  title={Label-Flipping Attacks in GNN-Based Federated Learning}, 
  year={2025},
  volume={12},
  number={2},
  pages={1357-1368},
  doi={10.1109/TNSE.2025.3528831}}

@inproceedings{Byzantine-robust-learning-on-heterogeneous-data-via-gradient-splitting,
author = {Liu, Yuchen and Chen, Chen and Lyu, Lingjuan and Wu, Fangzhao and Wu, Sai and Chen, Gang},
title = {Byzantine-robust learning on heterogeneous data via gradient splitting},
year = {2023},
publisher = {JMLR.org},
booktitle = {Proceedings of the 40th International Conference on Machine Learning},
articleno = {884},
numpages = {22},
location = {Honolulu, Hawaii, USA},
series = {ICML'23}
}

@article{JMLR:v26:24-1307,
  author  = {Jie Peng and Weiyu Li and Stefan Vlaski and Qing Ling},
  title   = {Mean Aggregator is More Robust than Robust Aggregators under Label Poisoning Attacks on Distributed Heterogeneous Data},
  journal = {Journal of Machine Learning Research},
  year    = {2025},
  volume  = {26},
  number  = {27},
  pages   = {1--51},
  url     = {http://jmlr.org/papers/v26/24-1307.html}
}

@inproceedings{Karimireddy2020ByzantineRobustLO,
  title={Byzantine-Robust Learning on Heterogeneous Datasets via Bucketing},
  author={Sai Praneeth Karimireddy and Lie He and Martin Jaggi},
  booktitle={International Conference on Learning Representations},
  year={2020},
  url={https://api.semanticscholar.org/CorpusID:238856649}
}

@article{DBLP:journals/corr/abs-1802-10116,
  author       = {Cong Xie and
                  Oluwasanmi Koyejo and
                  Indranil Gupta},
  title        = {Generalized Byzantine-tolerant {SGD}},
  journal      = {CoRR},
  volume       = {abs/1802.10116},
  year         = {2018},
  url          = {http://arxiv.org/abs/1802.10116},
  eprinttype    = {arXiv},
  eprint       = {1802.10116},
  bibsource    = {dblp computer science bibliography, https://dblp.org}
}

@InProceedings{pmlr-v139-radford21a,
  title = 	 {Learning Transferable Visual Models From Natural Language Supervision},
  author =       {Radford, Alec and Kim, Jong Wook and Hallacy, Chris and Ramesh, Aditya and Goh, Gabriel and Agarwal, Sandhini and Sastry, Girish and Askell, Amanda and Mishkin, Pamela and Clark, Jack and Krueger, Gretchen and Sutskever, Ilya},
  booktitle = 	 {Proceedings of the 38th International Conference on Machine Learning},
  pages = 	 {8748--8763},
  year = 	 {2021},
  editor = 	 {Meila, Marina and Zhang, Tong},
  volume = 	 {139},
  series = 	 {Proceedings of Machine Learning Research},
  month = 	 {18--24 Jul},
  publisher =    {PMLR},
  url = 	 {https://proceedings.mlr.press/v139/radford21a.html}
}

@misc{khosla2021supervisedcontrastivelearning,
      title={Supervised Contrastive Learning}, 
      author={Prannay Khosla and Piotr Teterwak and Chen Wang and Aaron Sarna and Yonglong Tian and Phillip Isola and Aaron Maschinot and Ce Liu and Dilip Krishnan},
      year={2021},
      eprint={2004.11362},
      archivePrefix={arXiv},
      primaryClass={cs.LG},
      url={https://arxiv.org/abs/2004.11362}, 
}

@article{Tian2020RethinkingFI,
  title={Rethinking Few-Shot Image Classification: a Good Embedding Is All You Need?},
  author={Yonglong Tian and Yue Wang and Dilip Krishnan and Joshua B. Tenenbaum and Phillip Isola},
  journal={ArXiv},
  year={2020},
  volume={abs/2003.11539},
  url={https://api.semanticscholar.org/CorpusID:214641252}
}

@InProceedings{pmlr-v119-chen20j,
  title = 	 {A Simple Framework for Contrastive Learning of Visual Representations},
  author =       {Chen, Ting and Kornblith, Simon and Norouzi, Mohammad and Hinton, Geoffrey},
  booktitle = 	 {Proceedings of the 37th International Conference on Machine Learning},
  pages = 	 {1597--1607},
  year = 	 {2020},
  editor = 	 {III, Hal Daumé and Singh, Aarti},
  volume = 	 {119},
  series = 	 {Proceedings of Machine Learning Research},
  month = 	 {13--18 Jul},
  publisher =    {PMLR},
  url = 	 {https://proceedings.mlr.press/v119/chen20j.html}
}
\bibliographystyle{tmlr}

\appendix

\section*{Appendix}
This appendix is organized as follows, section~\ref{appdx:exp_setup} details the datasets, data splits, target models, and training protocol, and provides the end-to-end pseudocode used in our experiments. 
We then justify in Section~\ref{appdx:discussion} the first focus on mean aggregation
and that our labels-only attack operates within valid gradient magnitudes, limiting the effectiveness of norm-based defenses. Subsequent sections discuss data distribution assumptions (our attack is distribution-agnostic and heterogeneity can further hinder robust aggregators) and clarify the threat model: we analyze an omniscient adversary for comparability and outline how a surrogate-based attacker can operate with partial knowledge.

\section{Dataset and Experimental Setup}
\label{appdx:exp_setup}

\paragraph{Datasets.}
We perform all experiments on MNIST10 and CIFAR10. Each image was scaled by its unit norm.








\paragraph{Implementation Details.} 
We train the models for 200 epochs using mini-batch SGD with a batch size of 64 and a learning rate of 0.001, using cross-entropy loss. At each epoch, the omniscient attacker observes the current model parameters and gradients, then flips the labels of a randomly assigned subset $K$ from the clean data pool accordingly. All results are averaged over six independent runs with different random seeds. The global training algorithm can be found below (Algorithm~\ref{alg:poisoned_training}).

\begin{algorithm}[h]
\caption{Full Training with Per-Round Label-Flipping Attack}
\label{alg:poisoned_training}
\begin{algorithmic}[1]
\REQUIRE Clean training set $D_{\mathrm{train}}$, model $h_{\theta}$, number of training rounds $T$, write-access fraction $k$, local flipping budget $b$, and functions:
    \begin{itemize}
        \item \texttt{sampleBatch}: samples a clean batch $D_t^H \subset D_{\mathrm{train}}$.
        \item \texttt{getWriteAccess}: selects the attacker's controlled subset $K_t^H \subset D_t^H$ with $|K_t^H| = \lfloor k |D_t^H| \rfloor$.
        \item \texttt{selectFlip}: applies Algorithm~\ref{algo:multiclass} to choose at most $\lfloor b |K_t^H| \rfloor$ labels to flip.
        \item \texttt{trainStep}: performs one optimization step on the submitted batch.
    \end{itemize}
\ENSURE Poison-trained model $h_{\theta_T}$
\STATE Initialize model parameters $\theta_0$
\FOR{$t \gets 0$ \textbf{to} $T-1$}
    \STATE $D_t^H \gets \texttt{sampleBatch}(D_{\mathrm{train}})$
    \STATE $K_t^H \gets \texttt{getWriteAccess}(D_t^H, k)$
    \STATE $H_t \gets D_t^H \setminus K_t^H$
    \STATE $K_t(\tilde y) \gets \texttt{selectFlip}(K_t^H, \theta_t, b)$ 
    \STATE $D_t(\tilde y) \gets H_t \cup K_t(\tilde y)$
    \STATE $\theta_{t+1} \gets \texttt{trainStep}(\theta_t, D_t(\tilde y))$
\ENDFOR
\STATE \textbf{return} $h_{\theta_T}$
\end{algorithmic}
\end{algorithm}

\aekedit{
\paragraph{Robust-aggregation and LIE comparison setup.}
For the robust-aggregation experiments in Figures~\ref{fig:robust_agg_cifar} and~\ref{fig:robust_agg_mnist}, we evaluate coordinate-wise median and trimmed mean. For trimmed mean, we remove the largest and smallest 10\% of submitted values in every coordinate. 

For the LIE baseline, each minibatch contains 1600 examples partitioned among \(n_w=100\) clients, with 16 examples per client. For a Byzantine fraction \(\rho\), we set \(m=\operatorname{round}(\rho n_w)\) malicious clients. Each malicious client submits
$
g_{\mathrm{mal}} = \mu_h - 3\sigma_h \operatorname{sign}(\mu_h),
$
where \(\mu_h\) and \(\sigma_h\) are the coordinate-wise mean and population standard deviation of the honest client gradients. We evaluate \(\rho\in\{0.05,0.10,\ldots,0.45\}\).

For label-flipping attacks, \(k\) is the attacker-controlled fraction of the data, \(b\) is the fraction of that subset whose labels are changed, and the plotted budget is \(kb\in\{0.05,0.10,\ldots,0.45\}\). 
}

\section{Discussion}
\label{appdx:discussion}
\subsection{How practical is the attack?}

In addition to our formal setup, allowing comparison between gradient attacks and label flipping, we note several realistic settings where label-only corruption can occur in production systems while possibly having access to knowledge about the model. This can be the case in, {\it e.g.},
\begin{itemize}
    \item compromised labeling tools: attackers who can modify annotation UIs, metadata, or labeling scripts can change labels stored in databases while leaving raw sensor data untouched,
    \item insider at a data provider: an employee with access to label fields (but not feature pipelines) can flip labels locally, 
    \item crowdsourced annotation manipulation: attackers controlling a subset of crowd workers or poisoning weakly-supervised labeling heuristics can flip a small fraction of labels across contributors,
    \item collaborative learning situations where participants are provided various forms of ``a global model'' in exchange for their participation with data as in, {\it e.g.}~~\cite{capitaine2024unravelling, ananthakrishnan2024delegating}.
\end{itemize}

\subsection{Why Logistic Regression (before deeper models)}
\label{how-deeper-models}
There are two reasons why we adopt logistic regression as a first-class test bed before moving to deeper models as we did: 
\emph{(i)  Analytic transparency.}  
The loss is convex and its gradient admits a
simple closed form, allowing us to derive our greedy attacks.
\emph{(ii)  Gateway to richer models.}  
Any classifier whose final layer is linear followed by a sigmoid/softmax
(e.g.\ multilayer networks, convolutional nets) reduces to logistic regression
if we treat the penultimate activations as “features.”
Consequently, at each training step our attack should apply to those
models by operating on the last‐layer logits.%
\footnote{Formally, let $f_\theta(X)=\sigma(W h_\phi(X))$ with inner network
$h_\phi$ fixed during one optimization step.  Setting
$\tilde X \!=\!h_\phi(X)$ recasts the update as a logistic
regression in $(\tilde X,y)$ with parameter $W$.}
This raises the question: Would such an attack still be greedy for any classifier?
One path to performing a label-flipping attack on a broader class of models could be the transition from logistic regression to neural networks. We perform experiments on a 2-layer vanilla MLP using the method described above. The results in Figure~\ref{fig:normalized-accuracy-vs-kb} show that the attack is remarkably efficient even for these deeper models.
 Another approach is to find a globally optimal label flipping attack algorithm. We assumed that the server was aggregating the gradients using the mean, and future work could study similar attacks on settings with different aggregation methods presented in ~\cite{
 baruch2019little, mhamdi2018hidden, 
 blanchard2017byzantinetolerant}. 

\subsection{Why mean aggregation still matters}
 \label{appdx:Why_mean_aggregation_still_matters}

 
 Mean aggregation remains the de facto baseline in both federated learning research and large-scale deployments as it is favored for its scalability, communication efficiency, and ease of implementation (~\cite{DBLP:conf/aistats/McMahanMRHA17}). As noted in~\cite{DBLP:journals/corr/Back_to_the_Drawing_Board}, production cross-device FL systems with thousands to billions of clients often employ plain averaging, and empirical evidence suggests that even the most basic, non-robust FL algorithms can be surprisingly resilient in realistic, low-compromise settings. In this context, showing that a constrained, label-flipping adversary can meaningfully degrade model performance under mean aggregation is already a strong and practically relevant finding.

Our focus on mean aggregation is therefore intentional: it establishes a baseline benchmark for the impact of constrained label-flipping attacks. If such an attack fails under mean aggregation, it is unlikely to pose a serious threat to more sophisticated robust aggregators; if it succeeds, it identifies a vulnerability that robust methods should be expected to address. Moreover, our attack is derived specifically for the mean operator, and different aggregation rules may require different attack formulations. Nevertheless, our experiments show that the same label-selection rule also transfers empirically to the tested robust aggregators, namely coordinate-wise median and trimmed mean. 

Finally, constrained label flipping interacts differently with defenses than arbitrary gradient manipulation. Since the attack changes only labels while leaving feature inputs untouched, the resulting updates are still induced by valid training examples and feasible labels. As a result, norm-based defenses such as clipping, which are often used in combination with mean aggregation~\cite{sun2019reallybackdoorfederatedlearning}, may fail to distinguish poisoned updates when their magnitudes remain within honest-looking ranges. This further motivates studying mean aggregation as a clean and practically relevant baseline before moving to more complex aggregation and defense mechanisms.

\subsection{Scaling affects both attack interpretation and detectability.}
Normalization affects both the strength and the detectability of the attack. Theoretically, the flip score contains inner products of the form
\[
\langle \Delta_t,x_i\rangle
=
\|\Delta_t\|\,\|x_i\|\cos(\phi_i).
\]
When features are normalized, the score is driven primarily by directional alignment. When features are unnormalized, high-norm points can dominate the ranking, making the optimized attack less purely directional. This can reduce the gap between optimized and random flipping in some settings.

Normalization also matters for detectability. In normalized settings, we empirically observe that the attacker’s per-client gradient norm is nearly indistinguishable from honest clients (see Table~\ref{tab:grad_norms_normalized} in Appendix~\ref{appdx:grad_norms}). Thus, defenses based only on gradient-norm clipping may fail to distinguish malicious updates from honest ones. 


\subsection{What if the attacker lacks omniscience} \label{appdx:On_the_threat_model}
In the case where the attacker lacks full knowledge~\cite{bouaziz2026winter}, for example consider a scenario where the attacker knows the architecture of the server’s model but not its current parameters. Then the attacker can train a local surrogate model using only the data available to them (i.e., the subset they control). This local model may not perfectly match the server’s model, but it may still approximate the decision boundary or the gradient dynamics well enough to inform a useful attack as done in, {\it e.g.}~\cite{mhamdi2018hidden, baruch2019little} or anticipate the dynamics in one-time attack as in~\cite{LablePoisoningIsAllYouNeed}.

Specifically, the attacker can simulate the label-flipping attack on their local model concurrently with the server’s training. By observing how different label flips affect the gradient direction or training loss on their surrogate, the attacker can estimate which flips are most likely to have the desired effect when sent to the server. \aekedit{ This strategy would probably reduce the attack's effectiveness given that less grounded information is available to the adversary, but remains as a viable option. 
}


\subsection{On the data distribution}
\label{appdx:data_dist}
Our experiments used a uniform i.i.d. data split across workers. All clients sampled from the same distribution and contributed equal-sized batches. However, our attack is distribution-agnostic: it does not assume i.i.d. data and optimizes flips online per iteration. In fact, heterogeneity generally makes many robust aggregation rules less reliable; non-IID gradients enlarge natural variability, which both weakens anomaly tests and can cause robust rules (e.g., trimming) to underperform FedAvg in certain regimes (~\cite{Byzantine-robust-learning-on-heterogeneous-data-via-gradient-splitting}).

As attacks are easier in heterogeneous settings, and harder in homogeneous ones (\cite{JMLR:v26:24-1307, Karimireddy2020ByzantineRobustLO}), our work highlights the power of label flipping even on i.i.d. (and thus less vulnerable) situations.

\section{Detectability and data normalization}
\label{appdx:grad_norms}
\begin{table}[hbt]
  \centering
  \caption{Normalized-data mean gradient-norm ratio (attacker mean gradient norm divided by honest mean gradient norm). Ratios extremely close to~1 indicate that, under standard dataset normalization, the attacker’s per-round gradient magnitudes are statistically indistinguishable from those of honest clients.}

  \label{tab:grad_norms_normalized}
  \begin{tabular}{l l r}
    \toprule
    \textbf{dataset} & \textbf{model} & \textbf{attack/honest norm ratio} \\
    \midrule
    cifar10  & logreg  & 1.004523 \\
    cifar10  & vanilla & 1.000400 \\
    mnist10  & logreg  & 1.028216 \\
    mnist10  & vanilla & 1.018608 \\
    \bottomrule
  \end{tabular}
\end{table}


In this section, we evaluate how detectable our label-flipping attack is under common feature-preprocessing pipelines, focusing on the gradient-norm statistics that many FL defenses monitor. Practical FL systems almost always apply some form of input normalization or scaling before training. Under these normalized settings, we empirically find that the attacker’s per-client gradient norms become nearly indistinguishable from those of honest clients (Table \ref{tab:grad_norms_normalized}). As a result, defenses that rely solely on gradient-norm thresholds or magnitude-based outlier filtering are ineffective. 

\aecomment{
\input{TMLR_v2_rebuttal}
}

\end{document}